\documentclass[a4paper,12pt]{article}
\pdfoutput=1
\usepackage{feynmp-auto,expdlist}
\usepackage{amsmath, amsfonts, amssymb,mathbbol}
\usepackage{graphicx}
\usepackage{enumerate,cancel}
\usepackage{hyperref}
\usepackage{latexsym}
\usepackage{dsfont}
\usepackage{hepnicenames}
\usepackage{enumerate}
\usepackage{soul}
\usepackage[normalem]{ulem}
\usepackage{comment}
\usepackage{units}

\newcommand{\gfootnote}[1]{\footnote{\tiny\color{gray}#1}}
\renewcommand{\gfootnote}[1]{}
\usepackage{soul}

\usepackage{mathrsfs,graphicx,rotating,amsmath,amsfonts,mathtools,booktabs,amssymb,wasysym}
\usepackage{hyperref}\usepackage{slashed}
\usepackage[nosort]{cite}
\usepackage[table,xcdraw,dvipsnames]{xcolor}
\usepackage{bm}
\usepackage{graphicx}
\usepackage{multirow,multicol}
\hypersetup{colorlinks,bookmarksopen,bookmarksnumbered,
	linkcolor=blus,pdfstartview=FitH,urlcolor=rossos,citecolor=verde}
\allowdisplaybreaks

\renewcommand{\[}{\left[}

\newcommand{\Lag}{\mathscr{L}}

\newcommand{\mio}[1]{}

\newcommand{\med}[1]{\langle #1\rangle}

\newcommand{\bpm}{\begin{pmatrix}}
\newcommand{\epm}{\end{pmatrix}}

\usepackage{mathrsfs}

\newcommand{\fig}[1]{~\ref{fig:#1}}
\newcommand{\sfrac}[2]{#1/#2}

\newcommand{\beqa}{\begin{equation}\begin{aligned}}
\newcommand{\eeqa}{\end{aligned}\end{equation}}

\allowdisplaybreaks
\usepackage{multicol}
\usepackage{color}
\definecolor{rosso}{cmyk}{0,1,1,0.4}
\definecolor{rossos}{cmyk}{0,1,1,0.55}
\definecolor{rossoc}{cmyk}{0,1,1,0.2}
\definecolor{blu}{cmyk}{1,1,0,0.3}
\definecolor{blus}{cmyk}{1,1,0,0.6}
\definecolor{bluc}{cmyk}{1,1,0,0.1}
\definecolor{verde}{cmyk}{0.92,0,0.59,0.25}
\definecolor{verdec}{cmyk}{0.92,0,0.59,0.15}
\definecolor{verdes}{cmyk}{0.92,0,0.59,0.4}

\newcommand{\bp}{\bar{M}_{\rm Pl}}
\newcommand{\eq}[1]{~{\rm (\ref{eq:#1})}}

\newcommand{\TeV}{\,{\rm TeV}}

\newcommand{\beq}{\begin{equation}}
\newcommand{\eeq}{\end{equation}}

\newcommand{\bea}{\begin{eqnarray}}
\newcommand{\eea}{\end{eqnarray}}
\newcommand{\be}{\begin{equation}}
\newcommand{\ee}{\end{equation}}
\font\tenrsfs=rsfs10 at 12pt
\font\sevenrsfs=rsfs7
\font\fiversfs=rsfs5
\newfam\rsfsfam
\textfont\rsfsfam=\tenrsfs
\scriptfont\rsfsfam=\sevenrsfs
\scriptscriptfont\rsfsfam=\fiversfs
\def\be#1\ee{\begin{equation}#1\end{equation}}
\def\bl#1\el{\begin{align}#1\end{align}}
\def\ba#1\ea{\begin{align*}#1\end{align*}}

\renewenvironment{thebibliography}[1]
{\begin{multicols}{2}[\section*{\refname}]%
		\@mkboth{\MakeUppercase\refname}{\MakeUppercase\refname}%
		\list{\@biblabel{\@arabic\c@enumiv}}%
		{\settowidth\labelwidth{\@biblabel{#1}}%
			\leftmargin\labelwidth
			\advance\leftmargin\labelsep
			\@openbib@code
			\usecounter{enumiv}%
			\let\p@enumiv\@empty
			\renewcommand\theenumiv{\@arabic\c@enumiv}}%
		\sloppy
		\clubpenalty4000
		\@clubpenalty \clubpenalty
		\widowpenalty4000%
		\sfcode`\.\@m}
	{\def\@noitemerr
		{\@latex@warning{Empty `thebibliography' environment}}%
		\endlist\end{multicols}}

\newcommand{\SU}{\,{\rm SU}}

\newcommand{\SO}{\,{\rm SO}}
\newcommand{\U}{\,{\rm U}}
\makeatletter
\font\ital=cmu10

\def\hhref#1{\href{http://arxiv.org/abs/#1}{arXiv:#1}}
\usepackage{xstring}
\newcommand{\hhrefq}[1]{\IfSubStr{#1}{:}{\href{http://inspirehep.net/search?ln=en&ln=en&p=#1&of=hb&action_search=Search&sf=&so=d&rm=&rg=25&sc=0}{InSpire:#1}}{\hhref{#1}}}

\def\art{\@ifnextchar[{\eart}{\oart}}
\def\eart[#1]#2#3#4#5#6{{\rm #2}, {\em #3 \bf #4} {\rm (#6) #5} ({\em #1})}
\def\article{\@ifnextchar[{\earticle}{\oarticle}}
\def\oarticle#1#2#3#4#5#6{{\rm #1}, {\ital `#6'}, {\rm #2 #3 (#5) #4}}
\def\earticle[#1]#2#3#4#5#6#7{{\rm #2}, {\ital `#7'}, {\rm #3 #4 (#6) #5}  [\hhrefq{#1}]}
\def\hepart[#1]#2{{\rm #2, \sl#1}}
\def\heparticle[#1]#2#3{#2, {\ital `#3'} [\hhrefq{#1}]}
\newcommand{\doi}[1]{\href{http://dx.doi.org/#1}{[link]}}

\newcommand{\hhrefqq}[1]{\IfBeginWith{#1}{10.}{\href{https://doi.org/#1}{doi:#1}}{\hhrefq{#1}}}

\renewenvironment{thebibliography}[1]
{\begin{multicols}{2}[\section*{\refname}]%
		\@mkboth{\MakeUppercase\refname}{\MakeUppercase\refname}%
		\list{\@biblabel{\@arabic\c@enumiv}}%
		{\settowidth\labelwidth{\@biblabel{#1}}%
			\leftmargin\labelwidth
			\advance\leftmargin\labelsep
			\@openbib@code
			\usecounter{enumiv}%
			\let\p@enumiv\@empty
			\renewcommand\theenumiv{\@arabic\c@enumiv}}%
		\sloppy
		\clubpenalty4000
		\@clubpenalty \clubpenalty
		\widowpenalty4000%
		\sfcode`\.\@m}
	{\renewcommand{\@noitemerr}
		{\@latex@warning{Empty `thebibliography' environment}}%
		\endlist\end{multicols}}

\newcommand{\eqnsystem}[1]{
	\renewcommand{\@eqnnum}{{\rm (\thealphaequation)}}
	\renewcommand{\@@eqncr}{\let\@tempa\relax \ifcase\@eqcnt \def\@tempa{& & &} \or
		\newcommand{\@tempa}{& &}\or \newcommand{\@tempa}{&}\fi\@tempa
		\if@eqnsw\@eqnnum\refstepcounter{alphaequation}\fi
		\global\@eqnswtrue\global\@eqcnt=0\cr}
	\refstepcounter{equation} \let\@currentlabel\theequation \def\@tempb{#1}
	\ifx\@tempb\empty\else\label{#1}\fi
	\refstepcounter{alphaequation}
	\let\@currentlabel\thealphaequation
	\global\@eqnswtrue\global\@eqcnt=0 \tabskip\@centering\let\\=\@eqncr
	$$\halign to \displaywidth\bgroup \@eqnsel\hskip\@centering
	$\displaystyle\tabskip\z@{##}$&\global\@eqcnt\@ne
	\hskip2\arraycolsep\hfil${##}$\hfil& \global\@eqcnt\tw@\hskip2\arraycolsep
	$\displaystyle\tabskip\z@{##}$\hfil
	\tabskip\@centering&\llap{##}\tabskip\z@\cr}
\def\endeqnsystem{\@@eqncr\egroup$$\global\@ignoretrue} \makeatother

\newcounter{alphaequation}[equation]
\renewcommand{\thealphaequation}{\theequation\hbox to
	0.6em{\hfil\alph{alphaequation}\hfil}}

\definecolor{Gray}{gray}{0.95}

\begin{document}

\begin{center}
\bigskip\bigskip
{\bf\Large\color{red!60!black} How particle physics stops ultra-relativistic \\ 
bubbles and shapes their gravitational waves}\\
\bigskip
{\bf Alessandro Strumia}  \\[2ex]
{\em Dipartimento di Fisica, Universit\`a di Pisa, Italia}\\
\bigskip
{\bf\color{blus} Abstract}
\begin{quote}
The gravitational-wave spectrum produced by ultra-relativistic bubbles in a first-order cosmological phase transition can be approximated in terms of their energy density as a function of bubble size.
Bubble collisions and post-collision classical evolution negligibly change it.
The dominant  energy loss is particle scattering, often in the Landau--Pomeranchuk--Migdal regime.
So the GW spectrum depends on the particle-physics coupling (interpolating between the `envelope' and `bulk flow' limits), on the interaction type (gauge, Yukawa, scalar quartic, \dots), and on whether particle number is conserved.
Finally, we consider broken global symmetries, which give a mixture of `elastic' and `inelastic' collisions.
\end{quote}
\end{center}
\thispagestyle{empty}

\tableofcontents

\section{Introduction}\label{sec:issue}
If a strongly supercooled first-order phase transition occurred in the early Universe, 
the motion of expanding bubbles could have generated a detectable gravitational-wave (GW) background~\cite{Witten:1984rs,Hogan:1986qda,Kosowsky:1991ua,2008.09136,2305.02357,1903.09642}. 
Strong supercooling can naturally arise in theories with classical scale invariance, where symmetry breaking is radiative~\cite{2307.04694,1604.05035,1704.01034,1704.04955,1805.01473,1809.11129}.
The mass $m$ of the scalar $\phi$ driving the phase transition can be well beyond the weak scale:
gravitational waves offer a unique probe of such high-energy physics.
However, the window is limited because GW
are sourced by the collective, anisotropic motion of energy rather than by the microscopic properties of individual particles.

We consider models where bubbles `run away', 
meaning that they reach ultra-relativistic Lorentz factors $\gamma\gg1$~\cite{BodekerMoore,1703.08215},
typically up to $\gamma m\sim \bp$.
Lattice simulations~\cite{1906.02588,1912.00997,2007.04967} can only reach moderate $\gamma$, in view of the huge hierarchy of scales:
from cosmological scales,  to the particle-physics scale $1/m$, down to
the Lorentz-contracted wall thickness $\sim 1/\gamma m$.
Nevertheless the ultra-relativistic limit leads to simplifications that allow a partial analytic understanding. 
We show that the GW spectrum $\Omega_{\rm GW}(\omega)$ 
is a sort of transform of $\sigma(R) $, the Lorentz-enhanced wall surface energy density as a function of the bubble radius $R \approx t$.

Ultra-relativistic bubble collisions initially proceed through nearly free passage,
with little change in the wall Lorentz factor during the overlap~\cite{GiblinHuiLimYang,AhlqvistEckerleGreene}.
Our main goal is thereby to compute $\sigma(R)$ after collisions.

Previous studies modelled the post-collision evolution using power laws inferred from classical two-bubble lattice simulations~\cite{2007.04967,2012.07826}, see also~\cite{2208.11697}.
In this approach, the decay of the peak of the gradient energy density~\cite{2007.04967} or of the energy momentum tensor
$T$~\cite{2012.07826}, $\max T\propto R^{-\xi}$, 
is converted into a thin-wall source factor $f \propto R^{-\xi-1}$ 
(corresponding, in our notation, to $\sigma \propto R^{-\xi}$)
by assuming that the wall thickness remains constant after collision.
The authors of \cite{2007.04967} claim that part of the energy spreads into the collided volume, making the peak decay faster,
and report $\Omega_{\rm GW}\propto\omega^3$ at low frequency for a real scalar, 
and $\Omega_{\rm GW}\propto\omega$ for a complex scalar breaking a global $\U(1)$, 
attributing the slower decay to the phase difference between bubbles, which propagates as a sharp phase wall.
Furthermore~\cite{2012.07826} claims $\Omega_{\rm GW}\propto\omega^{2.3}$ for a gauged $\U(1)$.
An exponential damping was assumed in~\cite{1707.03111} as a phenomenological parametrisation.

\smallskip

We argue that  this needs substantial revision.
The peak stress does not determine the GW source: longitudinal spreading can reduce it without suppressing the radially integrated anisotropic stress. What matters is the total energy density $\sigma(R)$.
We find that classical energy losses are negligible in the ultra-relativistic limit,
where the dominant effect is instead the scattering of the quanta in the wall on particles in the ambient plasma~\cite{2306.15555,2403.05615}.
Energy losses of high-energy particles in matter are a well-studied non-trivial 
field~\cite{BetheHeitler,LPM,hep-ph/9802442}, that we generalize to our context.
Different energy losses arise depending on the size and on the
type of the interaction: a scalar quartic, a Yukawa, or a gauge interaction.
Furthermore, interactions that conserve the number of wall particles lead to their energy degradation, rather than to their removal.
The dominant effect often is medium-induced splitting, the analogue of bremsstrahlung, in the Landau--Pomeranchuk--Migdal regime~\cite{LPM,BDMPS,AMY}.
Goldstone interactions have no generic additional suppression at energies above the scales where their low-energy effective description applies.

Depending on the interaction size, our GW results are intermediate between `envelope' and `bulk flow'.
We keep using the standard terminology, although it is now non-descriptive and even misleading:
`envelope' (very fast energy losses after bubble collision)~\cite{Kosowsky:1991ua,astro-ph/9211004,0806.1828};
`bulk flow' (negligible energy losses)~\cite{1707.03111,1712.06869};
`elastic' (bubble intersection in the false vacuum)~\cite{1906.02588,1912.00997,2308.13070};
`inelastic' (bubble intersection in the true vacuum).

\smallskip

The paper is organised as follows.
In section~\ref{sec:GW} we specialise the general GW formul\ae{} to thin ultra-relativistic walls,
expressing the spectrum in terms of the post-collision evolution of the wall energy density,
and showing how the infrared tail $\Omega_{\rm GW}\propto\omega$ of persistent walls turns into $\Omega_{\rm GW}\propto\omega^3$ below the inverse time $1/t_{\rm stop}$ after which the walls disappear.
In section~\ref{sec:spreading} we show that classical post-collision dynamics negligibly affects the wall energy.
In section~\ref{sec:pulsestopping} we compute energy losses due to scatterings of the wall quanta with the plasma,
mediated by Yukawa, quartic, gauge and gravitational interactions.
In section~\ref{sec:gamma(t)} we derive the resulting evolution of the wall energy density, summarised in table~\ref{tab:sigmaR}, and the GW spectra.
In section~\ref{sec:DVbrake} we discuss `elastic' collisions and theories with multiple vacua and (pseudo-)Goldstone bosons.
Conclusions are given in section~\ref{sec:concl}.


%
%

\section{Gravitational wave spectrum from bubble dynamics}\label{sec:GW}

In general, the GW spectrum is determined as
\beq  \frac{d\Omega_{\rm GW}}{d\ln\omega}=\frac{1}{\rho_{\rm tot}V}\int d\Omega\,
  \frac{dE}{d\Omega\,d \ln \omega},\qquad
  \frac{dE}{d\Omega\,d \ln \omega}
=2G\omega^3 \, |\widetilde T_{ij}^{\rm TT}(\omega, \omega \hat{\bm{k}})|^2
\eeq
by the light-cone value of the Fourier transform of
the transverse-traceless (TT) spatial stress:
\beq
\widetilde T_{\mu\nu}(\omega,\bm{k})
=\int\frac{d^4x}{2\pi}\,
e^{i(\omega t-\bm{k}\cdot\bm{x})}\,T_{\mu\nu}(x).
\label{eq:stress}
\eeq
The exponential in eq.\eq{stress} encodes in Fourier space the retarded-time integral. 
Here $E$ is the GW energy emitted from a large volume $V$, and $d\Omega$ is the solid angle of $\hat{\bm k}$.

\medskip

We consider a first-order phase transition approximated by the usual two parameters:
$\alpha=\Delta V/\rho_{\rm rad}$, the ratio between the released vacuum energy and the radiation energy density;
and $\beta$, the inverse duration of the transition, defined by a bubble nucleation rate per unit volume $\Gamma\propto e^{\beta t}$.
Assuming $\beta/H\gg1$, so that cosmological expansion is negligible during the bubble collision epoch~\cite{2511.15687},
the GW spectrum at production, normalised to the total energy density, is conventionally parameterised as~\cite{0806.1828}
\beq
\frac{d\Omega_{\rm GW}}{d\ln\omega}\bigg|_{\rm production}=\Big(\frac{H}{\beta}\Big)^2\Big(\frac{\kappa\,\alpha}{1+\alpha}\Big)^2\,\Delta(\omega/\beta),
\label{eq:OmegaDelta}
\eeq
where $H$ is the Hubble rate at the transition,
$\kappa$ is the fraction of the vacuum energy carried by the bubble walls,
and $\Delta$ is a dimensionless spectral function.
$\Delta$ also depends on the parameters of the post-collision dynamics, discussed in sections~\ref{sec:spreading}--\ref{sec:DVbrake}.
The spectral function in eq.\eq{OmegaDelta} is
\beq
\Delta(\omega/\beta) = \frac{3\beta^2\omega^3}{4\pi\kappa^2\Delta V^2\, V}\int d\Omega\,
|\widetilde T_{ij}^{\rm TT}(\omega, \omega \hat{\bm{k}})|^2.
\label{eq:Delta}
\eeq
For a scalar field $\phi$ one has $T_{ij}=\partial_i\phi\,\partial_j\phi-\delta_{ij}\Lag$:
the term proportional to $\delta_{ij}$, which contains the kinetic and potential energy, is removed by the TT projection,
so only spatial gradients source GW.

\subsection{GW sourced in the wall approximation}
A rigid stress distribution moving at constant velocity $\bm v$ with $v<c$ does not radiate,
since its Fourier transform in eq.\eq{stress}
is supported at $\omega=\bm k\cdot\bm v<|\bm k|$, away from the light-cone $\omega=|\bm k|$.
As only gradients contribute, GW are sourced by the bubble walls,
while the homogeneous regions inside and among bubbles contribute negligibly.

We assume that bubbles become ultra-relativistic, $\gamma\gg1$, before colliding.
At frequencies $\omega$ below the inverse of the Lorentz-contracted wall thickness,
the energy-momentum tensor of a bubble nucleated at $x_n^\mu=(t_n,\bm x_n)$ can be approximated as a thin wall moving at the speed of light,
\beq
 T_{\mu\nu}(x)\approx\sigma_n(R,\hat{\bm n})\,U_\mu U_\nu\,\delta\big(R-(t-t_n)\big),\qquad
 U^{\mu}=(1,\hat{\bm n}),\qquad \hat{\bm n}=\frac{\bm x-\bm x_n}{R},
 \label{eq:Twall}
\eeq
up to corrections of order $1/\gamma^2$.
Here $R=|\bm x-\bm x_n|$  is the bubble radius
and $\sigma_n(R,\hat{\bm n})$ is the energy per unit area of the element of the wall of bubble $n$ that moves along $\hat{\bm n}$,
when it is at radius $R$.
Before collision $\sigma_n=\gamma\sigma_0$ is isotropic.
After collision $\sigma_n$ depends on direction.

Inserting eq.\eq{Twall} into eq.\eq{stress}, the $\delta$-function fixes $t=t_n+R$ and leaves $d^3x=R^2\,dR\,d\Omega_{\hat{\bm n}}$.
On-shell, $\bm k=\omega\hat{\bm k}$, each wall element thereby contributes with phase $e^{i\omega(1-\mu)R}$, where $\mu=\hat{\bm k}\cdot\hat{\bm n}=\cos\theta$.
The TT projection of $U_iU_j=\hat n_i\hat n_j$ is
\beq
\Lambda_{ij,lm}\,\hat n_l\hat n_m=\frac{1-\mu^2}{2}\sum_{\lambda=+,\times}g_\lambda(\varphi)\,e^\lambda_{ij},\qquad
g_+=\cos2\varphi,\qquad g_\times=\sin2\varphi,
\eeq
where $\Lambda_{ij,lm}=P_{il}P_{jm}-\frac12P_{ij}P_{lm}$ with $P_{ij}=\delta_{ij}-\hat k_i\hat k_j$;
$\varphi$ is the azimuth of $\hat{\bm n}$ around $\hat{\bm k}$, measured from a unit vector $\bm e_1\perp\hat{\bm k}$;
$e^+=\bm e_1\bm e_1-\bm e_2\bm e_2$ and $e^\times=\bm e_1\bm e_2+\bm e_2\bm e_1$, with $\bm e_2=\hat{\bm k}\times\bm e_1$,
are normalised as $e^\lambda_{ij}e^{\lambda'}_{ij}=2\delta^{\lambda\lambda'}$.
Summing over all bubbles, the spectral function of eq.\eq{Delta} becomes
\beqa
\Delta(\omega/\beta)&=\frac{3\beta^2\omega^3}{32\pi^3\kappa^2\Delta V^2\,V}\int d\Omega_{\hat{\bm k}}\sum_{\lambda=+,\times}
\big|\mathcal{A}_\lambda(\omega,\hat{\bm k})\big|^2,\\
\mathcal{A}_\lambda(\omega,\hat{\bm k})&=\sum_n e^{i\omega(t_n-\hat{\bm k}\cdot\bm x_n)}\int d\Omega_{\hat{\bm n}}\,(1-\mu^2)\,g_\lambda(\varphi)
\int_0^{\infty}dR\,R^2\sigma_n(R,\hat{\bm n})\,e^{i\omega(1-\mu)R}.
 \label{eq:Deltawall}
\eeqa
The angular factors have simple interpretations:
\begin{itemize}
\item[-] The phase $e^{i\omega(1-\mu)R}$ becomes stationary for $\mu\to1$: a wall element moving along $\hat{\bm k}$ at the speed of light stays in phase with the wave it emits.
The TT factor $1-\mu^2$ forbids such forward emission, so the limit $v\to c$ is regular.
\item[-] The azimuthal factors $g_\lambda$ imply that a distribution of $\sigma_n$ axially symmetric around $\hat{\bm k}$ does not radiate.
In particular a single spherical bubble does not radiate: only the anisotropy induced by collisions contributes,
although the radial integral starts at nucleation, not at collision.
\end{itemize}
The GW spectrum is thereby a sort of transform of $R^2\sigma_n(R,\hat{\bm n})$. 
Different evolutions lead to different GW spectra.
We start from a simplifying case: walls that keep their energy
 after colliding at radius $R\approx R_c$, 
 until suddenly disappearing after a time $t_{\rm stop}$.
So $R^2\sigma_n$ is constant in the intermediate phase. 
Their contribution to the radial integral is proportional to
\beq
\int_{R_c}^{R_c+t_{\rm stop}}dR\,e^{i\omega(1-\mu)R}=e^{i\omega(1-\mu)R_c}\,\frac{e^{i\omega(1-\mu)t_{\rm stop}}-1}{i\omega(1-\mu)}.
\eeq
Two regimes arise at low frequencies $\omega R_c\ll1$:
\begin{itemize}
\item[-] For $\omega t_{\rm stop}\ll1$ the radial integral tends to $t_{\rm stop}$:
$\mathcal{A}_\lambda$ is regular for $\omega\to0$ and $\Delta\propto\omega^3$.

\item[-] For $\omega t_{\rm stop}\gg1$ one gets 
$\mathcal{A}_\lambda\propto 1/\omega$ and $\Delta\propto\omega$.
Indeed the rapidly oscillating term averages out in the angular integral,
and the radial integral scales as $1/[\omega(1-\mu)]$.
The angular factor $(1-\mu^2)/(1-\mu)=1+\mu$ is finite.
This is Weinberg's zero-frequency limit of gravitational bremsstrahlung~\cite{Weinberg:1965nx},
which applies to any process that permanently redistributes energy into an anisotropic flux of relativistic matter.

\end{itemize}
The `bulk flow' limit $t_{\rm stop}\to\infty$ thereby gives $\Delta\propto\omega$ down to arbitrarily low frequencies,
while the `envelope' limit $t_{\rm stop}\lesssim R_c$ gives $\Delta\propto\omega^3$ below the peak.

\medskip

One final technical comment.
The GW integrals in eq.\eq{stress} can be evaluated with two alternative techniques:
\begin{enumerate}[a)]
\item Via numerical simulations~\cite{1712.06869}, which are made easy by approximating bubbles as thin walls.
The frequency range is limited at low frequencies by the box size, and at high frequencies by the integration grid spacing.

\item By analytically computing the ensemble-averaged $\med{T_{\mu\nu}(x)T_{\mu'\nu'}(x')}$.
This gives up to 7 nested integrals~\cite{1605.01403,1707.03111}, while in the previous method
$T_{\mu\nu}$ is simply computed and squared.
This method allows one to extract analytically the limits of large and small $\omega$.
\end{enumerate}
We use method a).

\section{Post-collision classical bubble wall dynamics}\label{sec:spreading}
Before collision, energy conservation with negligible friction implies that $\gamma = R \, \Delta V/3\sigma_0 = R/R_0$
grows linearly (runaway walls), where $R_0 = 3\sigma_0/\Delta V$ is the {thin-wall} radius of the critical {O(4)} bubble.
In models with gauge interactions 
 friction can instead saturate the growth at constant $\gamma \sim  \Delta V/g^2 M_V  T^3$~\cite{BodekerMoore,1703.08215}.
At fixed geometry and evolution,
uniformly reducing $\gamma$ reduces the GW spectrum quadratically,
$\Omega_{\rm GW}\propto \gamma^2$;
the energy dissipated by friction can nevertheless produce sound waves or other plasma motions that act as a different GW source.

\smallskip

In the ultra-relativistic limit colliding walls initially pass through each other, with their fields superposing, up to $1/\gamma^2$ corrections~\cite{GiblinHuiLimYang,AhlqvistEckerleGreene}.
 After collision, two main opposite limiting cases are considered in the literature:
\begin{itemize}
\item `{\bf Envelope}' approximation~\cite{Kosowsky:1991ua,astro-ph/9211004,astro-ph/9310044,0806.1828,1605.01403}.
If walls disappear rapidly after collision, the source has a finite duration. 
As we just discussed, this leads to  $\Omega_{\rm GW}\propto\omega^3$ at low frequencies.
This approximation was adopted in the early literature, assuming that walls are stopped by their
collisions. But this is not the case in the ultra-relativistic limit.
Nevertheless, we will show that post-collision particle dynamics can have a similar effect.


\item `{\bf Bulk flow}' approximation~\cite{1707.03111,1712.06869}. Neglecting interactions, walls keep expanding freely after collision:
energy conservation implies the geometric decrease $\sigma\propto 1/R^2$,
compensated by the area growing as $R^2$.
This corresponds to gravitational memory from persistent anisotropic outgoing energy.
A persistent source behaves in time as a step function switched on when it appeared,
so its Fourier transform has a $1/\omega$ tail at low $\omega$, resulting in $\Omega_{\rm GW}\propto \omega$.
Intuitively, the low-$\omega$ enhancement arises because the freely expanding shells become large.
\end{itemize}
We here want to understand how the surface energy density of a bubble wall evolves after the collision.
We will show how particle physics leads to intermediate situations:
if the bubbles eventually disappear after a time
$t_{\rm stop}$, the spectrum returns to $\omega^3$ at $\omega t_{\rm stop}\ll1$.
We consider a generic real scalar $\phi$ with mass $m$ and potential $V(\phi)$.
Without loss of generality we can assume that the true vacuum is at $\phi_{\rm true}=0$.
The real scalar could be the radial mode of a complex scalar, 
or its light pseudo-Goldstone angular mode.

The subsequent sections consider `inelastic' collisions, where the true vacuum is restored inside bubbles.
`Elastic' collisions, where the false vacuum is restored in bubble intersections, will be discussed in section~\ref{sec:DVbrake}.

\subsection{Classical field evolution after `inelastic' collisions}\label{sec:class}
In this section we show that classical field evolution negligibly contributes to  energy losses of ultra-relativistic walls,
clarifying some misconceptions in the literature.

As a preliminary note, we recall that one spherical bubble in coordinates 
$t$ and $R=\sqrt{x^2+y^2+z^2}$ is described by the wave equation
\beq \label{eq:onebubble}
\frac{\partial^2\phi}{\partial t^2}-\frac{\partial^2\phi}{\partial R^2}-\frac{2}{R}\frac{\partial\phi}{\partial R}=-V'(\phi).\eeq
Next, we consider the post-collision evolution of two spherical bubble walls colliding at large Lorentz factor $\gamma$,
so that each bubble is approximately O(3,1) symmetric around its centre.
If plasma friction is negligible (its effect will be considered in the next sections), 
the post-collision physics has an O(2,1) symmetry.
Different parts of the two bubbles collide at different times and thereby with different Lorentz factors, 
but this is compensated by a transverse boost.
The field then depends only on two coordinates.
In the frame where the bubbles nucleate simultaneously at $z=\pm d$,
O$(2,1)$ acts on $(t,x,y)$.
The coordinate $\tau=\sqrt{t^2-r^2}$ with $r^2=x^2+y^2$ makes this symmetry explicit: the field depends only
on $(\tau,z)$, and its equation is
\begin{equation}
\label{eq:pde}
\frac{\partial^2\phi}{\partial \tau^2}
+\frac{2}{\tau}\,\frac{\partial\phi}{\partial \tau}
-\frac{\partial^2\phi}{\partial z^2}
=-V'(\phi).
\end{equation}
A single O(3,1)-symmetric bubble centred at $z=\mp d$ corresponds to a solution that depends only on $\tau^2-(z\pm d)^2$.
This extra geometry encoded in eq.\eq{pde} reduces to the simpler eq.\eq{onebubble}
once the relativistic front of each wave at $\tau \approx |z|+d$
gets causally disconnected from what happens behind.


\smallskip

The substitution $\Phi=\tau \phi$ simplifies eq.\eq{pde} into
\beq
 \frac{\partial^2\Phi}{\partial \tau^2}
 -\frac{\partial^2\Phi}{\partial z^2}
 =  - \tau V'( \frac{\Phi}{\tau})=  -m^2\Phi - \frac{\lambda}{\tau^2} \Phi^3
 \qquad \hbox{for}\qquad V (\phi) = \frac{m^2}{2}\phi^2+ \frac{\lambda}{4}\phi^4.
 \label{eq:linearphase}
\eeq
In view of the spherical geometry the interaction term becomes less relevant as the bubbles grow:
at late times the field intensity dilutes and one can approximate the potential as a mass term.
So in section~\ref{postcollisionwall} we consider a simple free field, and discuss interactions in section~\ref{sec:scalarself}.

\subsubsection{Free field: dispersion}\label{postcollisionwall}
For a massless free field the general solution is the sum of left-moving and
right-moving waves with geometric dilution,
\beq \phi(\tau,z)=\frac{F_R(z-\tau)+F_L(z+\tau)}{\tau}.\eeq
For a massive free field, the general solution is
\beq
\phi(\tau,z)=\frac1\tau\int_0^\infty\frac{dk}{2\pi}
\left[
A_R(k)e^{ikz-iE_k\tau}
+A_L(k)e^{-ikz-iE_k\tau}
+\text{c.c.}
\right],
\qquad E_k=\sqrt{k^2+m^2}.
\eeq
The evolution of the region inside the bubbles where $z$ does not matter is obtained by setting $k=0$:
the bulk phase responds on a time scale $1/m$, and oscillates around the minimum while diluting.
For $m=0$, instead, there is a constant memory mode $\Phi = {\rm cte}$,
similarly to an electromagnetic pulse that leaves a potential difference between after and before its passage.

\begin{figure}[t]
$$\includegraphics[width=0.68\textwidth]{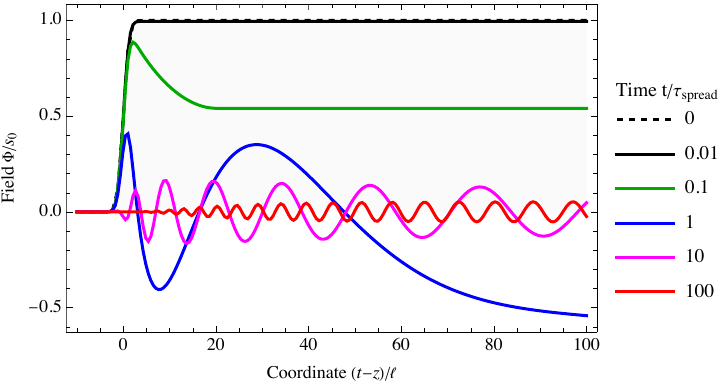}$$
\caption{\label{fig:AdvancingFront}\em 
Field profile of the front of a massive free bubble wall $\Phi(t,z)$ as it advances:
its length increases as $t$;
its energy density decreases as $1/t$;
its shape changes from a wall to a pulse.}
\end{figure}

Fig.\fig{AdvancingFront} shows an explicit numerical example of how a massive wall front evolves:
it appears to initially deposit some energy; 
next it spreads and becomes a pulse, rather than a wall.
While the field profile is complicated, the key physics is trivial:
{\em each Fourier mode moves forever at its speed}.
For a quantum field, this means that its quanta dilute due to the spherical geometry, keeping their energy.

The energy that appears to be deposited initially is the low-frequency tail of
non-relativistic modes, that carry a small fraction $\sim 1/\gamma$ of the total energy.
Once they remain behind, deposition ends and the wall has become a pulse.

The time needed for this process can be estimated in terms of the
characteristic momentum and width
$k\sim\Delta k \sim 1/\ell \sim  {\gamma/\ell_0}$, where $\ell_0$ is the wall thickness in its rest frame.
The  velocity varies across the packet by
$\Delta v\sim m^2\ell^2$. So its width changes appreciably at a time after collision
\beq
 \tau_{\rm spread}\equiv {\frac{\gamma}{m^2\ell_0}},
 \label{eq:spread}
\eeq
that reduces to $\gamma/m$ for $\ell_0\sim1/m$ 
and is longer for a light pseudo-Goldstone.
The same estimate is obtained from energy conservation.
Initially, advancing the front by a distance $dt$ leaves behind
a layer with potential-energy density
$\Delta V\approx m^2\phi^2/2$, costing an energy
$\Delta V\,dt$ per unit area.
The initial front energy per unit area is
$\sigma\sim\phi^2/\ell$, so substantial energy transfer from the front into the relativistic wake behind it occurs at
$
{\tau\sim\sfrac{\sigma}{\Delta V}\sim \tau_{\rm spread}}$.
This is a redistribution within the pulse, not an energy loss: a freely dispersing pulse conserves its energy.

Once dispersion becomes important, the packet spreads over a length proportional to $\tau$.
So the front $\Phi$ energy density decreases as $1/\tau$ and the $\Phi$ field amplitude as $1/\sqrt{\tau}$. 
The energy density of $\phi$ decreases as $\tau^{-3}$.
These results are confirmed by a stationary-phase approximation.
The spread is irrelevant for GW production, as it remains small compared to the relevant GW wavelengths.\footnote{Previous works interpreted this spreading as an energy loss.
The wall-to-pulse transition is visible in the numerics of section 6.1 of~\cite{2403.05615}, where it is misinterpreted
as a permanent energy loss in eq.~(160).
In~\cite{2007.04967} the faster decrease of the peak gradient energy density was attributed to energy spreading into the collided volume,
while~\cite{2012.07826} argued that gradients remain localised, $L\ll R$.
Both then assumed a constant post-collision wall thickness $L$, converting the decrease of the peak of $\rho_{\rm grad}$ or $T_{zz}$ into an energy loss.
However the spread energy remains in a relativistic wake just behind the front, and $L\ll R$ does not imply a constant $L$.
Indeed dispersion explains the $\max\rho_{\rm grad}\propto t^{-3}$ found in~\cite{2007.04967} for a real scalar, 
compared to $t^{-2}$ for the massless phase wall of a global $\U(1)$:
the extra $1/t$ is the growth of $L$, so that both cases conserve energy, $\sigma\propto R^{-2}$.}

\subsubsection{Scalar self-interactions}\label{sec:scalarself}
The discussion above simply means that a post-collision free wall deposits
a negligible fraction of its energy.
The same conclusion holds when scalar self-interactions are taken into account.
Interactions can transfer energy from large-$k$ modes to modes with lower $k$,
which then remain behind the pulse.
However, for an ultra-relativistic wall this effect is again suppressed by $1/\gamma$,
because both the energy $E$ and the momentum $p$ are conserved.
An ultra-relativistic wall has $E\approx p$, so it cannot leave behind a substantial
amount of energy with small momentum.
The energy $\Delta E$ left behind with velocity $v$ is capped as
$\Delta E\,(1-v)\le E-p$.
Initially $E-p$ is dominated by the non-relativistic modes with $k\sim m$, so it is at most of order $E/\gamma$.
Numerical simulations up to $\gamma\sim100$ confirm this, and find that the energy
initially deposited behind is comparable to what is computed in the free case.
The subsequent energy deposition is even smaller, for two reasons.
First,  once these soft modes have remained behind, the pulse has a much smaller $E-p$,
as each relativistic mode contributes only $m^2/2k$.
Second, at later times the spherical geometry suppresses the interactions:
in eq.\eq{linearphase} the nonlinear terms in $\tau V'(\Phi/\tau)$ decrease as $1/\tau$ (cubic) and $1/\tau^2$ (quartic).

The overall conclusion is that the seemingly complicated
energy losses found by numerically solving the classical equation of
motion for moderate Lorentz factors can be neglected in the ultra-relativistic limit.

\section{Energy loss from scatterings on the plasma}\label{sec:pulsestopping}
The previous section showed that a bubble wall quickly evolves into a pulse of ultra-relativistic particles that spread spherically.
They can get stopped by interactions with ambient matter at temperature $T$.
Before being reached by the pulses, the plasma has only been heated by friction on the wall that crossed it.
So its temperature is $T = T_{\rm nuc}\,[1+(1-\kappa)\alpha]^{1/4}$
if the wall dissipated a fraction $1-\kappa$ of the vacuum energy.
This temperature lies between the nucleation temperature $T_{\rm nuc}$
and the reheating temperature $T_{\rm reh}=T_{\rm nuc}(1+\alpha)^{1/4}$ 
reached when all the vacuum energy is converted into radiation.
It can be much lower than $T_{\rm reh}$ in strongly supercooled transitions with runaway walls, 
where $\alpha\gg1$ and $1-\kappa\ll1$: bubble interiors are then nearly empty,
until the pulses themselves heat the plasma.
In the following we consider a generic $T$.
The total energy of walls is dominated by the hardest quanta with energy $E_c\sim \gamma_c m\gg T$.
The spherical geometry dilutes the quanta while keeping their energy.

The medium pressure on a pre-collision wall is usually computed as 
$P\sim\sum_i\Delta m_i^2T^2$~\cite{BodekerMoore}
for a bubble that expands in the false vacuum, 
giving extra mass $\Delta m$ to plasma particles.
This effect cancels when the wall becomes a pulse, which has the same masses on both sides.
Furthermore, the $\Delta m$ effect is an interaction with soft momenta of the wall, 
which carry a small fraction of the total energy; this is why this $P$ is not $\gamma$-enhanced.

In the post-collision phase, energy losses are instead similar
 to energy losses of ultra-relativistic particles in matter.

\subsection{Hard scatterings from renormalizable couplings}\label{sec:1/s}
Cross sections become small at ultra-high energy, $\hat\sigma \sim 1/\hat{s}$ where $\hat{s}$ is the invariant 
scattering energy.
An exception are cross sections mediated by a long-range Coulomb-like interaction, which can remain constant,
giving effects discussed later.
The relevant integrated cross section is
\beq \hat\sigma_{\rm tr}=\int d\hat t\,\frac{d\hat\sigma}{d\hat t}\frac{\min(-\hat t,-\hat u)}{\hat s}\eeq
where the weight follows the leading final-state particle.
This $\hat\sigma_{\rm tr}\lesssim 1/\hat{s}$ remains small even in the presence
of a long-range enhancement,
$d\hat\sigma/d\hat t  \propto 1/(M^2-\hat t)^2$.
The energy-loss rate is $dE/dt=-\Gamma_E E$, where the fractional rate $\Gamma_E$ is
\beq
\Gamma_E=\sum_p g_p\int\frac{d^3q}{(2\pi)^3}f_p(q)\,(1-\cos\vartheta)\,\hat\sigma_{p,\rm tr}(\hat s).
\eeq
The sum runs over the bath particles $p$, with $g_p$ internal degrees of freedom, momentum $q$
and distribution $f_p$; $\hat\sigma_p$ is averaged over their internal states;
$\vartheta$ is the angle between the momenta of the energetic quantum and of the bath particle
in the plasma frame, so that $1-\cos\vartheta$ is the relative velocity and $\hat s=2Eq(1-\cos\vartheta)$.
For $\hat\sigma_{p,\rm tr}=c_p/\hat s$ the factor $1-\cos\vartheta$ cancels out, leaving
\beq
\Gamma_E= k\,\frac{T^2}{E},\qquad
k=\sum_p\frac{g_p c_p}{24}
\begin{cases}
1 & \hbox{for a boson $p$},\\
1/2 & \hbox{for a fermion $p$}.
\end{cases}
\eeq
We assume that scattered bath particles thermalise;
this can induce fluid motion that sources extra gravitational waves, as discussed at the end of section~\ref{sec:gamma(t)}.
Each quantum  loses energy at the constant rate $dE/dt\approx-kT^2$
so that it stops at $t_{\rm stop} = E_c/k T^2$ if $E_c$ is the initial energy.
This is macroscopic, $t_{\rm stop} \sim (m/T)^2/k\beta$, 
for a runaway bubble with $\gamma_c \sim R_c/R_0\sim m/\beta$ (for $R_0\sim 1/m$).

\begin{figure}[t]
$$\includegraphics[width=\textwidth]{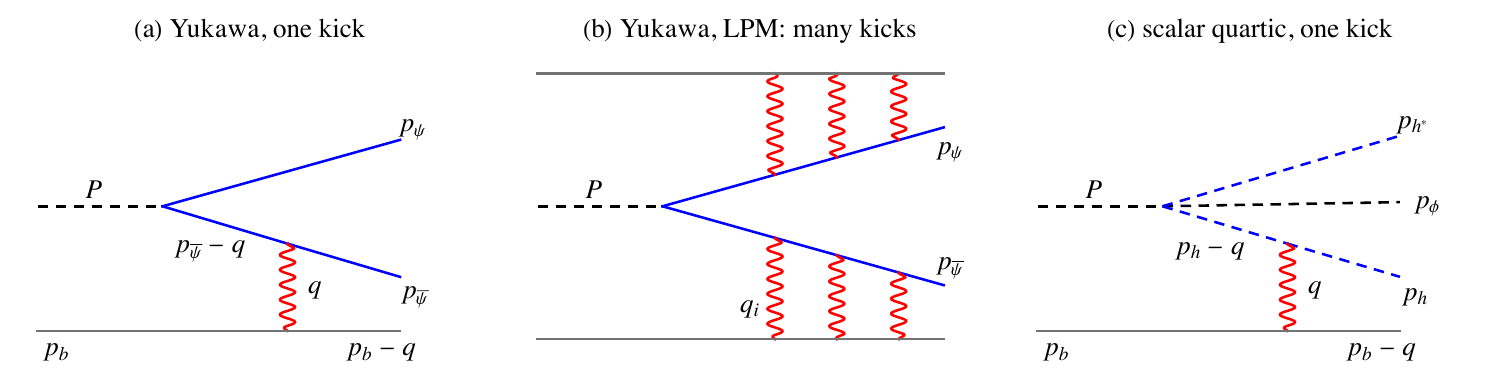}$$
\begin{center}
\caption{\em\label{fig:FeynmanLPMdiagrams} Feynman diagrams relevant for energy losses.}
\end{center}
\end{figure}

\subsection{Yukawa couplings}\label{sec:Yukawa}
We start by considering the possibility that the scalar $\phi$ has a 
Yukawa coupling $y\, \phi\bar\psi\psi/\sqrt{2}$ to a Dirac fermion $\psi$
that has extra interactions (for example, being charged under a gauge group),
and acquires the mass $m_\psi=yf/\sqrt2$ in the true vacuum. 
If $m>2m_\psi$ the dimensionless Yukawas can induce the  $\phi\to\psi\bar\psi$ decay
with rate $\Gamma (\phi\to\psi\bar\psi) \sim (m/E) \Gamma_{\rm rest} $, 
Lorentz suppressed at large energy $E$, just like the scatterings of section~\ref{sec:1/s}.
As discussed in the previous section, such effects lead to slow energy losses.

\medskip

A more important effect is $\phi p\to \psi\bar\psi p$ where $p$ is some bath particle with typical energy $T$,
which interacts with $\psi$ via $t$-channel exchange of a vector with gauge coupling $\alpha$:
$p$ can be any particle charged under the gauge group (including the vectors themselves, if the group is non-Abelian).
The SM plasma contains $\SU(3)_c\otimes\SU(2)_L\otimes\U(1)_Y$ vectors that interact with quarks, leptons and Higgs bosons.
For example, KSVZ axion models can give this physics with $\phi$ 
the PQ-breaking scalar, $\psi$ a quark and $p$ a gluon.

The vector provides the crucial new physics:
its Coulomb-like interaction favours soft momentum transfer, while nearly collinear kinematics allows the intermediate virtual fermion $\psi_{\rm vir}$ to remain close to its mass shell.
The $\phi\to\psi\bar\psi$ conversion can therefore proceed through a soft kick from the plasma, 
without requiring a hard momentum transfer of order $\sqrt{\hat s}$.
The left panel of fig.\fig{FeynmanLPMdiagrams} shows the relevant Feynman diagram
in the single-vector-exchange approximation:
 the rate is set by the soft-kick rate $\Gamma_{\rm kick} \sim \alpha T$ times the splitting probability proportional to $y^2$, giving $\Gamma(\phi p\to\psi\bar\psi p)\sim y^2\alpha T$.
This represents an $E/T$ enhancement in energy scaling relative to the naive $\Gamma\propto T^3/\hat s\sim T^2/E$ estimate.
This single-kick enhancement was first noticed by Bethe and Heitler in $e Z \to e Z\gamma$~\cite{BetheHeitler}
who explained the observed $dE/dx \approx - b E$ energy loss in matter.

However this enhancement only holds up to some energy; 
a softer $dE/dx \propto -\sqrt{E}$ applies at higher energies.
In this regime, relevant for our problem, the $E/T$ enhancement is partially lost because 
the conversion into a nearly collinear $\psi\bar\psi$ pair takes a finite formation time, 
$t_{\rm form}\sim E/Q^2$,  where $Q^2$ denotes the characteristic virtuality of the intermediate fermion.
Soft gauge scatterings occur at a rate $\Gamma_{\rm kick}\sim\alpha T$, 
each transferring a typical transverse momentum squared $\alpha T^2$.
The total is $Q^2\sim N\alpha T^2$ and the number $N$ of kicks is
\beq
N\sim\Gamma_{\rm kick}t_{\rm form}
\sim\alpha T \frac{E}{N\alpha T^2},
\qquad\hbox{hence}\qquad
N\sim\sqrt{\frac{E}{T}}.
\eeq
Thus, although the small initial virtuality permits a long formation time, the accumulated kicks increase the virtuality and limit that time self-consistently.
There is effectively one splitting opportunity per formation time, rather than one per kick, so
$\Gamma(\phi p\to \psi\bar\psi p) \sim\sfrac{y^2}{t_{\rm form}}
\sim\sfrac{y^2\alpha T}{N}$.
The rate is only enhanced by $(E/T)/N\sim\sqrt{E/T}$.
This effect is given by the middle Feynman diagram in fig.\fig{FeynmanLPMdiagrams} and
is known as the Landau--Pomeranchuk--Migdal effect, after the authors who understood it for high-energy electrons in matter~\cite{LPM}.
It was later confirmed experimentally~\cite{LPMexp} and extended to QCD~\cite{BDMPS}, thermal field theory~\cite{AMY}, and cosmology~\cite{1312.3097}.
Our interaction rate is thereby estimated as
\beq \Gamma (\phi p \to\psi\bar\psi p) \approx c  y^2 \,\alpha T\sqrt{\frac{T}{E}}
\eeq
where $c$ is an order unity coefficient and $\alpha=g^2/4\pi$ is the coupling of the gauge vector exchanged in the soft kicks. 
It is given by $c\approx0.5\sqrt{\ln(1/\alpha_3)}$ if $\psi$ is a quark and the vector is a gluon,  
having taken into account that the Yukawa coupling $y$ gives a flat splitting function.
In general $c\approx 0.1\, N_\psi\sqrt{C_\psi \ln(1/\alpha)}\,m_D/gT$, 
where $N_\psi$ is the dimension of the gauge representation of $\psi$, $C_\psi$ its quadratic Casimir,
{and} $m_D$ {the} Debye mass that screens soft gauge interactions in the plasma, $m_D^2=g^2T^2(N_c/3+N_f/6+N_s/6)$ for SU($N_c$) with $N_f$ Dirac fermions and $N_s$ complex scalars in the fundamental~\cite{Weldon:1982aq,AMY,0810.1026}.
This enhancement operates if fermions are light enough that their mass does not dominate the virtuality 
provided by the kicks, $m_\psi^2\ll N\alpha T^2$, which is satisfied by the hardest quanta even for fermions that get mass $\sim y\med{\phi} \sim y T$.

Then, the stopping time of a post-collision wall is
\beq
 t_{\rm stop} \approx {\frac{1}{c\, y^2 \alpha}} \frac{E_c}{T^2} \sqrt{\frac{T}{E_c}}.
\label{eq:tstopYukawa}
\eeq
Large enough gauge and Yukawa couplings can lead to fast stopping:
$t_{\rm stop}\lesssim R_c$ for
\beq
c\,\alpha y^2 \gtrsim 3~10^{-6} \sqrt{\frac{E_c}{\bp}\frac{T}{\TeV}}\,\frac{\beta}{100H}
\eeq
having used $TR_c\approx2~10^{13}\,(\TeV/T)(100H/\beta)$ for $R_c=(8\pi)^{1/3}/\beta$ and $g_*\approx 106.75$.

This $t_{\rm stop}$ describes the conversion of the scalar, while the GW source ends only once the daughters
have showered and isotropized: the subsequent gauge shower, with rate $\sim\alpha^2T\sqrt{T/E}$,
is faster than the conversion if $c\,y^2\lesssim\alpha$.
In this case the pulse is converted into showers that degrade their energy down to $\sim T$ and thermalise close to the collision,
so that the post-collision scalar source effectively ends at collision, as assumed in the `envelope' approximation.
The momentum carried by the pulse is deposited into the plasma, where it can start fluid motion~\cite{2403.05615},
which can source gravitational waves for a longer time (see the end of section~\ref{sec:gamma(t)}).

\subsection{Scalar quartic couplings}\label{sec:quartic}
A mixed quartic $\lambda_{\phi h}\phi^2h^2/4$ to a real bath scalar $h$  gives
$\hat\sigma_{\rm tr}(\phi h \to \phi h)=c/\hat s$ with $c=\lambda_{\phi h}^2/64\pi$ for $\hat s,|\hat t|\gg m^2$.
Similarly, if $\phi$ itself is in the bath, the self-quartic $\lambda_\phi \phi^4/4$ gives
$c=9\lambda_\phi^2/32\pi$. 
The associated cubic couplings and exchange diagrams give effects suppressed by $m^2/\hat s$.
These $2\to 2$ cross sections are suppressed at large energy, so
the resulting stopping time of a runaway bubble is macroscopic 
\beq t_{\rm stop}=\frac{E_c}{kT^2}\approx\frac{1536\pi}{N_h\lambda_{\phi h}^2+18\lambda_\phi^2}\frac{\gamma_c m}{T^2}.\eeq
This is the dominant effect if $h$ is a gauge singlet.

\smallskip

A larger effect arises if $h$ is gauge-charged (for example, if it is the Higgs doublet with $N_h=4$): 
soft gauge kicks induce the
$1\to3$ splitting $\phi\to \phi hh^*$ in the LPM regime discussed in the previous section.
To see this, consider the diagram in the right panel of fig.\fig{FeynmanLPMdiagrams}.
In terms of the momenta defined in the figure,
the final $\phi hh^*$ system has invariant mass $Q^2=(p_\phi+p_h+p_{h^*})^2=(P+q)^2\approx 2P\cdot q$.
If $h$ carries a fraction $z$ of the energy, $p_h\approx zP$,  
the virtual $h$ has a small virtuality
$(p_h-q)^2-m_h^2\approx-2p_h\cdot q\approx-zQ^2$,
while the gauge vertex $\propto p_h$ gives $z$ times the eikonal elastic amplitude $\mathscr{A}_{\rm el}\approx 2g^2\hat s/\hat t$.\footnote{No such enhancement arises if $h$ only has Yukawa or quartic interactions with the bath:
only $t$-channel vector exchange with amplitude $\propto1/\hat t$ gives the frequent soft kicks needed for LPM.}
The factors of $z$ cancel, so $|\mathscr{A}|^2\approx\lambda_{\phi h}^2|\mathscr{A}_{\rm el}|^2/Q^4$.
Factorizing the phase space as $d\Phi_4=d\Phi_2\,(dQ^2/2\pi)\,d\Phi_3$, with $\Phi_3=Q^2/256\pi^3$ for massless particles, gives
\beq d\sigma(\phi p\to \phi hh^* p)\approx d\sigma_{\rm el} \frac{\lambda_{\phi h}^2}{512\pi^4}\frac{dQ^2}{Q^2}.\eeq
One elastic kick thereby produces the $\phi hh^*$ state with a probability per logarithmic interval of $Q^2$ independent of $Q$, 
like a $1\to2$ splitting with dimensionless coupling.
Multiple kicks again give the LPM result:
the virtual $h$ lives for $\sim E/Q^2$, so that $N\sim\sqrt{E/T}$ kicks act coherently as one.
The diagram with the kick on $h^*$ has opposite charge, cancelling the amplitude for $q\to 0$ as needed for a neutral $\phi$;
this affects order-one factors.
So 
\beq \Gamma(\phi\to \phi hh^*)\approx c\lambda_{\phi h}^2\,\alpha T\sqrt{T/E}\eeq 
enhanced by $\sim\alpha\sqrt{E/T}$ with respect to the $2\to2$ rate.
A rough  estimate gives $c\sim 10^{-4}N_h \sqrt{C_h\ln(1/\alpha)}\,m_D/2gT$, where $C_h$ is the quadratic Casimir of $h$,
that is $c\sim10^{-3}$ for the Higgs.
 Since each splitting transfers an order-one energy fraction, the fractional energy-loss rate $\Gamma_E$ is of the same order.
After symmetry breaking the cubic $\lambda_{\phi h}\med{\phi}\,\phi h^2/2$ allows $\phi\to hh$, with rate suppressed as $1/E$.
Unlike in the Yukawa case, the final state contains a $\phi$ that keeps a fraction $x\sim1/2$ of the energy and splits again:
the pulse is degraded rather than depleted, as discussed in section~\ref{sec:gamma(t)}.

\subsection{Gauge interactions}\label{sec:pulsegauge}
Novel interactions arise if the scalar $\phi$ itself is charged under gauge interactions.
As a result, bubbles reach a terminal Lorentz factor before collision, if it is smaller than the run-away value $R_c/R_0$.
Indeed, in the pre-collision phase each bubble is subject to a friction $P\propto \gamma$ 
that limits its Lorentz factor to $\gamma_c \sim \Delta V/g^2 M_V  T^3$~\cite{1703.08215,2306.15555}.
The friction  comes from transition radiation of the vectors that acquire mass  from the phase transition.
In our previous language, it can be seen as a $ \phi p \to  V p$ scattering on bath particles $p$;
the cross section $\hat\sigma$ is constant due to $t$-channel vector exchange.
This process converts $\phi\to V$ without degrading energy, 
so most energy is carried by a shell of vectors with energies $\sim\gamma M_V $.

In these models most vacuum energy gets converted into the motion of vectors, rather than into the motion of $\phi$ walls.
As a result, individual quanta no longer have a runaway energy $E$. 
Their cross sections are less suppressed by $E$.
Unless the gauge coupling is tiny, such vectors do not free stream;\footnote{In addition to the
$2\to2$ and $3\to2$ interactions considered in~\cite{2403.05615}, 
vectors undergo medium-induced $2\to3$ splittings in the LPM regime,
with $\Gamma\sim\alpha^2T_{\rm nuc}\sqrt{T_{\rm nuc}/E}$ rate which is usually dominant,
further reducing the free-streaming regions compared to what was computed in~\cite{2403.05615}.
See~\cite{1312.3097} for a discussion of LPM vector scatterings in a different cosmological context.}
rather they push acoustic waves in the fluid, that can source significant GW.
This is a well-studied topic unrelated to the stream of particles studied in this work.
So, we will not consider models with a gauge-charged $\phi$.

\subsection{Gravitational interactions}
Gravitational energy losses are irreducible, but negligible:
they do not prevent a cosmologically long $t_{\rm stop}$, for the following reasons.
First, scatterings on the bath have a low invariant energy, $\hat s\sim ET\ll\bp^2$, even for quanta with $E\sim\bp$.
Graviton exchange mediates a Coulomb-like 
$\hat\sigma_{\rm tr}\sim \hat{s}/\bp^4$, leading to $\Gamma_E R_c\sim  (EH/\bp^2)(H/\beta)\ll1$.
Second, soft gravitational kicks do not induce LPM splittings:
since gravity couples to energy, all collinear particles are deflected by the same angle,
so that their relative transverse momenta (the dipole that drives splittings in the gauge case) are unchanged,
and only suppressed tidal effects remain.
Third, the emission of gravitons has no collinear enhancement, and thereby no LPM enhancement:
their coherent emission is the GW signal itself, which carries a small fraction $\sim(H/\beta)^2$ of the energy.

\section{Evolution of the energy density of the bubble wall}\label{sec:gamma(t)}
We are now in a position to approximate the surface energy density $\sigma$ of a post-collision bubble, as dictated by
energy balance 
\beq
\frac{d(\sigma R^2)}{dR}\approx -R^2\, P(R)
\eeq
in the light of our understanding of the dominant pressure.
In the limit of negligible $P$, namely $t_{\rm stop}\gg R_c$, collided walls keep propagating as thin shells with $\sigma\propto1/R^2$,
recovering the `bulk flow' approximation~\cite{1707.03111,1712.06869};
in the opposite limit of large $P$, namely $t_{\rm stop}\ll R_c$, the scalar source ends at collision,
recovering the `envelope' approximation.
Depending on the energy loss, intermediate cases arise: 
a finite $t_{\rm stop}$ cuts the $\Omega_{\rm GW}\propto\omega$ infrared tail of the bulk flow 
at $\omega\sim1/t_{\rm stop}$, below which $\Omega_{\rm GW}\propto\omega^3$.

\begin{table}[t]
\begin{center}
\renewcommand{\arraystretch}{1.5}
\begin{tabular}{lccc}
\toprule
Energy-loss mechanism & $\Gamma_E \propto$ & monochromatic pulse & flat spectrum pulse\\
\midrule
1) Hard degradation & $ T^2/E$ & $1-\rho$ & $1-\rho+\rho\ln\rho$\\
2) Hard removal & $ T^2/E_c$ & $e^{-\rho}$ & $E_2(\rho)$\\
3) LPM degradation & $ T^{3/2}/\sqrt{E}$ & $(1-\rho)^2$ & $1-4\rho+3\rho^2-2\rho^2\ln\rho$\\
4) LPM removal & $ T^{3/2}/\sqrt{E_c}$ & $e^{-\rho}$ & $2E_3(\rho)$\\
\bottomrule
\end{tabular}
\caption{\em\label{tab:sigmaR} Evolution of the pulse energy density,
$\sigma(R)=\sigma_c(R_c/R)^2\,F(\rho)$ with $\rho=(R-R_c)/t_{\rm stop}$,
for the four cases discussed in the text.
The function $F$ is shown for a monochromatic pulse with energy $E_c$
and for a pulse with flat initial energy spectrum up to $E_c$.
Here $E_n(\rho)=\int_1^\infty du\,e^{-\rho u}/u^n$ is the generalized exponential integral.
The functions are illustrated in fig.\fig{Lorentz}. 
The degradation functions vanish for $\rho\ge1$. These results assume continuous energy losses.
}
\end{center}
\end{table}

\medskip

We consider `inelastic' collisions of bubbles whose intersection transits to the true vacuum
(the `elastic' case is discussed in section~\ref{sec:DVbrake}).
As discussed in section~\ref{sec:spreading},  we can neglect dispersion and field non-linearities and focus on the particle interactions of section~\ref{sec:pulsestopping}.
The pulse is a shell of $n$ quanta per unit area, each with energy $E=\gamma m$, so that $\sigma=nE$.
Free propagation conserves $nR^2$ and $E$, so that $\sigma R^2$ is constant in the absence of interactions:
the spherical dilution $\sigma\propto1/R^2$ is compensated by the area.
Each quantum loses energy at the rate $\Gamma_E(E)\,E$, so that the pressure is $P=\sigma\,\Gamma_E(E)$ and,
using $t=R-R_c$ for the ultra-relativistic pulse,
\beq
\frac{d\ln(\sigma R^2)}{dR}=-\Gamma_E(E).
\eeq
Qualitatively different behaviours arise, mainly depending
on their energy dependence and size and
on whether interactions conserve the number of wall particles.
We here derive simple expressions for a monochromatic pulse with initial energy at collision $E_c$.
\begin{itemize}
\item[1)]  Hard scatterings mediated by renormalizable interactions
(section~\ref{sec:1/s}) give a fractional
energy-loss rate $\Gamma_E=kT^2/E$.
If these scatterings conserve the number of pulse particles,
and the energy transferred to the bath thermalises rapidly,
the continuous-loss approximation at constant $T$ gives
\beq
\sigma(R)\approx\sigma_c\left(\frac{R_c}{R}\right)^2
\left[1-\frac{R-R_c}{t_{\rm stop}}\right],
\qquad
t_{\rm stop}=\frac{E_c}{kT^2}.
\eeq
Stochastic losses spread the Lorentz factors of the beam particles, so we wrote
the evolution of the total pulse energy density $\sigma = n\med{E}$, where $n$ is the surface number density of pulse particles.
The stopping time can be compared to the collision radius, finding
\beq
\frac{t_{\rm stop}}{R_c}\approx \frac{1.2}{k}\,\frac{E_c}{\bp}\,\frac{\beta}{H},
\eeq
independently of $T$. Perturbative couplings contribute as $k\lesssim 1$,
so that this energy loss is typically mildly slow.


\begin{figure}[t]
$$\includegraphics[width=0.65\textwidth]{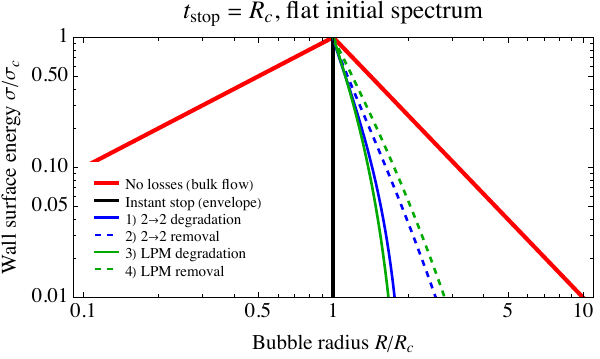}$$
\caption{\label{fig:Lorentz}\em 
Evolution of the wall energy density of
runaway bubbles  for the energy losses listed in table~\ref{tab:sigmaR}
assuming $t_{\rm stop}=R_c$,
together with the two limiting cases of very fast (black) and negligibly slow (red) energy losses.}
\end{figure}

\begin{figure}[p]
$$\includegraphics[width=0.45\textwidth]{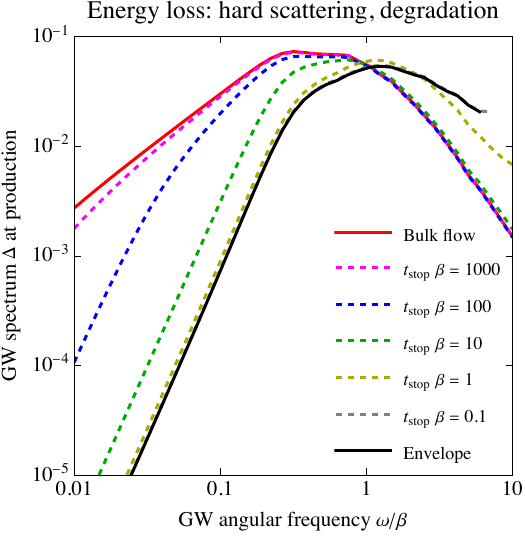}\qquad
\includegraphics[width=0.45\textwidth]{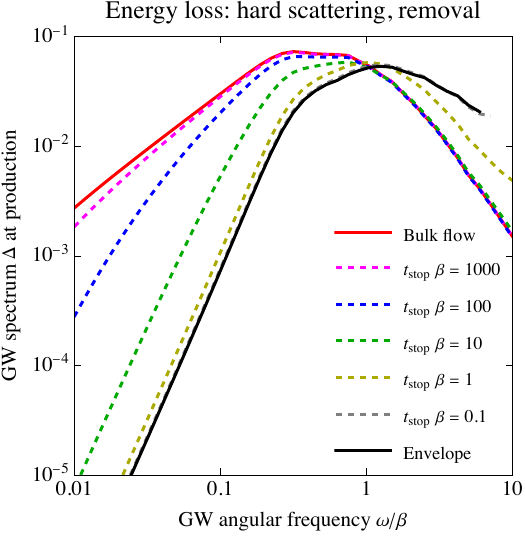}$$
$$\includegraphics[width=0.45\textwidth]{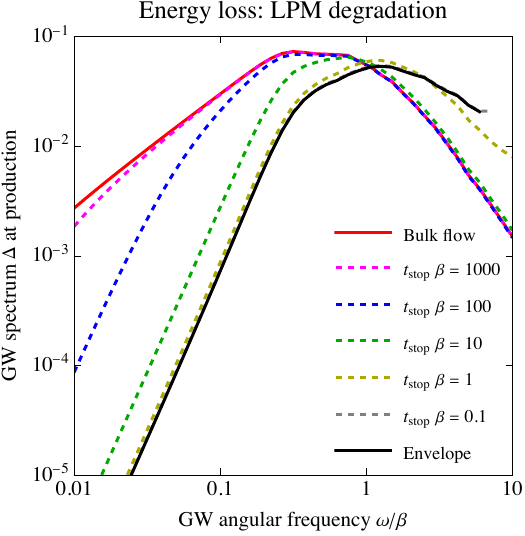}\qquad
\includegraphics[width=0.45\textwidth]{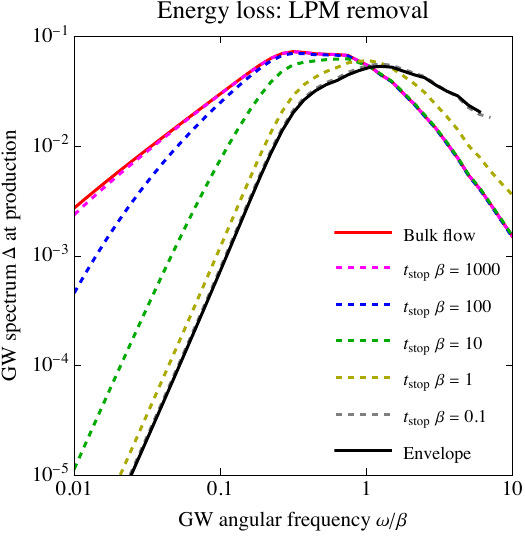}$$
\caption{\label{fig:GWSpectrumBubbleCollision}\em 
GW energy spectra generated by ultra-relativistic bubble walls 
with post-collision energy losses dominated by hard scatterings (upper row)
or Landau--Pomeranchuk--Migdal scatterings (lower row) with the cosmological plasma,
for quartic interactions that conserve particle number (left column),
or Yukawa interactions that remove particles (right column). 
The couplings dictate the value of the stopping time $t_{\rm stop}$.
Depending on the value of $t_{\rm stop} \beta$
(where $\beta \gg H$ is the inverse duration of the phase transition)
the GW spectrum ranges from the `envelope' to the `bulk flow' limit.
A gauge charged $\phi$ is not considered, as it typically does not lead to run-away bubbles.}
\end{figure}

\item[2)]
If instead each hard scattering removes a particle from the pulse,
and its products thermalise rapidly, the surviving particles retain
their initial energy $E_c$.
The fractional pulse-energy loss rate is therefore constant,
$\Gamma_E=kT^2/E_c$, rather than increasing as
$kT^2/E(R)$ when individual particles cool. This gives
\beq
\sigma(R)=\sigma_c\left(\frac{R_c}{R}\right)^2
e^{-(R-R_c)/t_{\rm stop}},
\qquad
t_{\rm stop}=\frac{E_c}{kT^2}.
\eeq
The effect is again typically slow: the pulse persists over distances
significantly larger than the collision radius.

\item[3)] Medium-induced splittings in the LPM regime give a larger
fractional energy-loss rate, less suppressed at high energy,
$\Gamma_E(E) \equiv kT^{3/2}/\sqrt{E}$ as in $\phi p \to \phi hh^* p$.
Assuming that scattering products rapidly thermalise gives
\beq
\sigma(R)\approx\sigma_c\left(\frac{R_c}{R}\right)^2
\left(1-\frac{R-R_c}{t_{\rm stop}}\right)^2,
\qquad
t_{\rm stop}=\frac{2}{\Gamma_E(E_c)}.
\eeq
Depending on the values of couplings this $t_{\rm stop}$ can be comparable to or smaller than $R_c$.

\item[4)] Finally, LPM splittings can remove particles from the pulse, as in $\phi p\to\psi\bar\psi p$.
If the products rapidly thermalise, the surviving particles retain the initial $\phi$ energy $E_c$,
so that the fractional pulse-energy loss rate is constant, $\Gamma_E=kT^{3/2}/\sqrt{E_c}$. This gives
again an exponential effect
\beq
\sigma(R)=\sigma_c\left(\frac{R_c}{R}\right)^2
e^{-(R-R_c)/t_{\rm stop}},
\qquad
t_{\rm stop}=\frac{1}{\Gamma_E(E_c)}=\frac{\sqrt{E_c}}{kT^{3/2}},
\eeq
but now $t_{\rm stop}$ can be smaller than $R_c$.\footnote{If the LPM suppression were not present, 
namely in the Bethe--Heitler regime valid only for $E\lesssim E_{\rm LPM}\sim T$,
the single-kick rate $\Gamma_{\rm BH}\sim c\,y^2\alpha T$ would be energy-independent.
Both degradation and removal would then give an exponential,
as for electromagnetic showers in matter, with microscopic $t_{\rm stop}\sim1/\Gamma_{\rm BH}$.}
\end{itemize}
The above results assume a monochromatic pulse with energy $E_c$.
Assuming instead a pulse with flat initial energy spectrum up to $E_c$, as typical of bubble walls,
gives the results summarised in table~\ref{tab:sigmaR} and computed by performing a convolution. 
Order-unity factors change as softer quanta stop earlier.

The above $t_{\rm stop}(T)$ assumes a constant plasma temperature $T$.
In strongly supercooled transitions the pulses instead initially hit a plasma much colder than $T_{\rm reh}$,
which is heated by the energy they themselves deposit.
Energy losses are then initially slow, and accelerate as the plasma approaches $T_{\rm reh}$.
Since $\Gamma_E \propto T^p$ with moderate $p=2$ ($3/2$) for hard (LPM) losses, this changes $t_{\rm stop}$ only by an order-one factor:
\beq t_{\rm stop} \approx  4t_{\rm stop}(T_{\rm reh})\int_{T_{\rm in}}^{T_{\rm reh}}\frac{dT}{T}
\left(\frac{T_{\rm reh}}{T}\right)^{p-4}
\simeq \frac{t_{\rm stop}(T_{\rm reh})}{1-p/4},\eeq
independently of the initial temperature $T_{\rm in}$, as for $p<4$ the integral converges and is dominated by $T\sim T_{\rm reh}$.

We normalized $t_{\rm stop}$ such that
in all cases the scalar pulse stops being a relevant GW source at $R\approx R_c+t_{\rm stop}$.
If $t_{\rm stop}\ll R_c$ the scalar source ends at collision, as assumed in the `envelope' approximation;
if $t_{\rm stop}\gg R_c$ the pulse propagates freely over many bubble sizes, 
so that the `bulk flow' approximation can hold down to low frequencies.
In both cases the energy lost by the pulse goes into collinear daughters 
within angles $\lesssim\sqrt{T/E}$ that retain its anisotropic stress.
The source is removed only once the cascade has degraded the energy down to $\sim T$ and isotropized it,
possibly leaving fluid motion, discussed below.
If a weakly interacting particle species appears along the cascade, it keeps moving.

Fig.\fig{Lorentz} illustrates the $\sigma(R)$ behaviours, and
fig.\fig{GWSpectrumBubbleCollision} shows the corresponding GW spectra.\footnote{The spectra are computed from eq.\eq{Deltawall} summing over many bubbles, doing analytically the integrals over $R$.}
For very long $t_{\rm stop}\gtrsim1/H$ one needs to take Hubble expansion into account~\cite{2511.15687}.
We neglected the response of the plasma to the energy and momentum deposited by the wall quanta, discussed next.

\subsection{Gravitational waves from fluid pushed by wall quanta}
Since the plasma thermalises locally on time scales much shorter than $R_c$, 
the deposition acts on it as a force density that drives collective fluid motion, 
such as sound waves or, for large deposited energy, relativistic fluid shells and shocks,
which source an extra GW component~\cite{2208.11697,1905.00899,2403.05615}.
Its size is controlled by the bulk velocity $v$ acquired by the fluid.
Momentum is conserved, but shared with the plasma:
since the anisotropic stress is roughly momentum density times velocity, $T_{ij}\approx \pi_i v_j$,
it is reduced by $v$ with respect to free-streaming wall quanta, which have $v=1$.
If the deposited energy is small compared to the plasma enthalpy $w$, then $v\approx \pi/w\ll 1$ and the stress is suppressed,
although the longer duration of sound waves partially compensates.
If instead the deposited energy dominates the local plasma,
the fluid moves relativistically and the stress is not reduced~\cite{1905.00899,2208.11697}.
Energy and momentum conservation imply that this happens for
\beq
\frac{t_{\rm stop}}{R_c}\lesssim\frac{\alpha_{\rm amb}}{4},\qquad
\alpha_{\rm amb}\equiv\frac{\Delta V}{\rho_{\rm rad}(T)}=\frac{\alpha}{1+(1-\kappa)\alpha},
\eeq
where $\rho_{\rm rad}(T)$ is the energy density of the plasma hit by the pulses, 
with temperature $T$ given in section~\ref{sec:pulsestopping}.
In this case, the relativistic motion of the fluid can generate a GW signal comparable to the wall contribution.
Rapid stopping therefore recovers the envelope limit for the wall source,
but not necessarily for the total GW signal.
Long-lived acoustic sources can also exhibit an intermediate
$\Omega_{\rm GW}\propto\omega$ regime~\cite{2308.12916,2308.12943},
extending down to their inverse lifetime rather than to $1/t_{\rm stop}$.


\begin{figure}[t]
$$\includegraphics[width=0.48\textwidth]{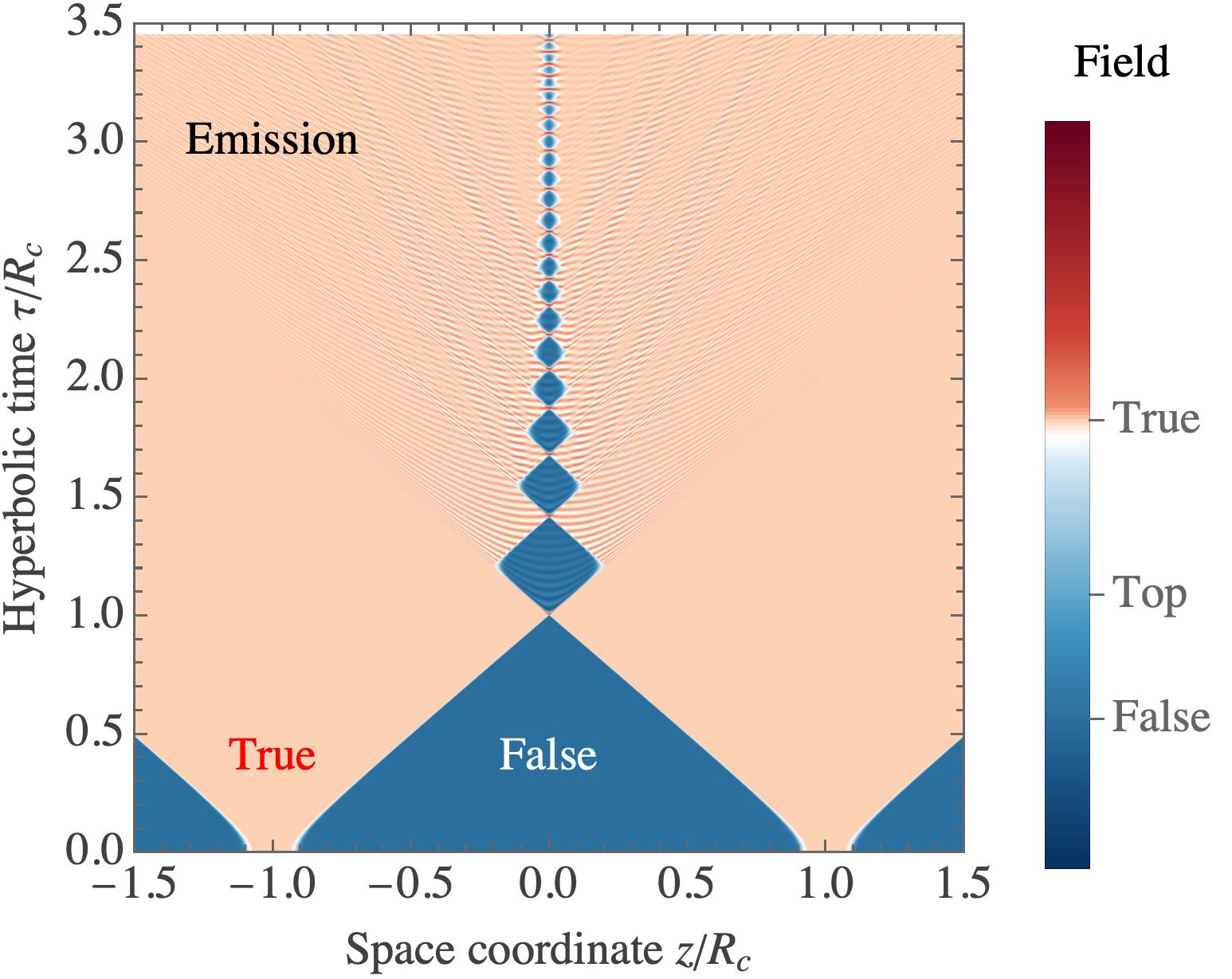}
\qquad
\includegraphics[width=0.4\textwidth]{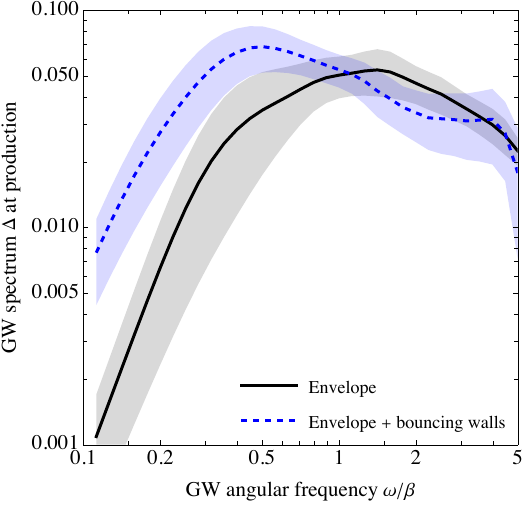}
$$
\caption{\label{fig:O21ElasticBounces}\em {\bf `Elastic' collision}.
We solve eq.\eq{pde} in the potential of~\cite{2308.13070} with $V_{\rm peak}/V_{\rm false}\approx1.7$,
obtaining an `elastic' collision of two bubbles with wall Lorentz factor $\gamma_c\approx11$. 
Left: the scalar field, as function of the hyperbolic time $\tau$ and 
of the coordinate $z$ along the collision axis, both in units of the bubble radius at collision $R_c$.
The walls repeatedly turn back and re-collide, with a decreasing amplitude.
Right: spectrum of gravitational waves.
To emphasize the effect of bouncing walls
we assumed that the radiated energy is quickly stopped,
and showed the correction with respect to the `envelope' approximation.
In the opposite limit GWs acquire a tail $\Omega_{\rm GW}\propto\omega$ at low frequency.
}
\end{figure}

\section{`Elastic' collisions and multiple vacua}\label{sec:DVbrake}
\label{sec:elastic}
So far we considered `inelastic' collisions, where the true vacuum is restored inside the intersection of two bubbles.
During free passage the field inside the intersection jumps to $\phi\approx2\phi_{\rm true}-\phi_{\rm false}$.
Depending on the potential, the subsequent evolution can restore the false vacuum, rather than the true vacuum.
This case is misleadingly dubbed `elastic', because in numerical results such as fig.\fig{O21ElasticBounces}a
it superficially looks as if the walls bounced back,
while they actually passed through each other.
The same vacuum energy difference $\Delta V$ that accelerated the bubbles before the collisions decelerates them
after the collision,
as the intersection region is left in the false vacuum, until the walls turn back and re-collide.
The walls perform a few re-collisions, damped by multiple effects:
\begin{enumerate}
\item[1)] At each re-collision an order-unity $\gamma$-independent fraction $\wp$ of the energy is radiated
away along the directions of the original walls.
Our numerical simulations find $\wp\approx(20-30)\%$, growing with $V_{\rm peak}/V_{\rm false}$.
This radiation sources GW similarly to walls after `inelastic' collisions.

\item[2)] The transverse expansion of the collision region dilutes the wall energy density during the expanding phase of the cycle,
\beq
\sigma \approx \left(\frac{R_c}{R}\right)^2\left[(1-\wp)\sigma_c-\frac{\Delta V}{3}\frac{R^3-R_c^3}{R_c^2}\right],
\eeq
and this is not compensated by an opposite effect during the contracting phase;
particles with $E\sim \gamma m$ are emitted in both phases.\footnote{Indeed, after collision the system is O(2,1) symmetric, with metric $ds^2=d\tau^2-dz^2-\tau^2dH_2^2$, where $dH_2$ is the unit hyperbolic plane.
A thin wall at $z=\pm Z(\tau)$ is described by the Lagrangian per unit comoving transverse area
$\Lag=-\sigma_0\tau^2\sqrt{1-\dot Z^2}-\Delta V\,\tau^2 Z$.
Its momentum per unit proper area is $\sigma_0\gamma v\approx\sigma$ for an ultra-relativistic wall, so that the equation of motion gives
$d(\tau^2\sigma)/d\tau\approx\mp\Delta V\,\tau^2$, where the upper sign holds while the walls move apart.
The proper area of each wall element grows as $\tau^2$ irrespective of the sign of $\dot Z$,
as in an expanding (Milne) universe, giving a Hubble-like friction $-2\sigma/\tau$ in both phases.
This is not an energy loss: the energy is carried by the outer parts of the collision region, which move transversally at nearly the speed of light.}

\item[3)] Energy losses on the plasma, as discussed in section~\ref{sec:pulsestopping}.
\end{enumerate}
Furthermore, bubbles of true vacuum can nucleate inside bubble intersections.

Fig.\fig{O21ElasticBounces} shows a numerical example: the walls turn back and re-collide a few times.
To isolate the effect of the re-collisions, 
we neglected GWs sourced by the outgoing radiation, assuming that it is promptly stopped by the plasma.
Fig.\fig{O21ElasticBounces}b shows that re-collisions mildly modify the GW spectrum around its peak at $\omega\sim \beta$.

\subsection{Phase transition breaking a global symmetry} \label{sec:freepassage}
We next explore whether the GW spectrum changes in theories
where the phase transition breaks an (approximate) global symmetry $G\to H$, giving light pseudo-Goldstones.

A first effect is that bubbles can nucleate along different directions in field space.
For a complex scalar $S$ charged under $G=\U(1)$ with false vacuum at $S=0$,
the initial field inside the overlap of two ultra-relativistic bubbles is
\beq
S_{\rm in} \approx S_1+S_2
 =\frac{f}{\sqrt2} \left(e^{i\theta_1}+e^{i\theta_2}\right)=\sqrt2 f \cos\frac{\Delta\theta}{2}\,e^{i(\theta_1+\theta_2)/2},
 \qquad \Delta\theta=\theta_1-\theta_2.
\eeq
For a potential barrier at $\sqrt2|S|=v_{\rm peak}(T)$, two cases arise:
\begin{itemize}
\item Overlaps with $|\cos(\Delta \theta/2)| > v_{\rm peak}/2f$ behave as in the single-field case:
they evolve towards the true vacuum, as in `inelastic' collisions,
unless they overshoot and return to the false vacuum as we discussed above.

\item A new undershoot mechanism implies that 
overlaps with $|\cos(\Delta \theta/2)| < v_{\rm peak}/2f$ start on the false side of the potential barrier
and evolve towards the false vacuum, as in `elastic' collisions.
\end{itemize}
These models thereby give a mixture of `elastic' and `inelastic' collisions,
with relative amounts that depend on the breaking pattern $G\to H$ and on the potential.
So the GW spectrum is intermediate between the two limiting cases illustrated in fig.\fig{O21ElasticBounces}.

A second effect is that the wall energy is shared between the radial and the Goldstone mode.
During free passage the field across the front entering bubble 1 moves along the straight chord from $S_1$ to $S_{\rm in}$, 
so that its gradient is parallel to $S_2$.
At each point $S$ along the chord the Goldstone fraction is $\sin^2\psi$, where $\psi$ is the angle between the chord and $S$.
Weighting each point of the chord by the wall energy density there,
which for a linear wall profile is uniform along the chord, 
and using that the chord subtends an angle $\Delta\theta/2$ at the origin, gives a Goldstone fraction
\beq F\approx\frac{\Delta\theta}{2}\sin\Delta\theta. \eeq
Assuming that bubbles nucleate equally in all field directions, the average Goldstone fraction is
$\med{F}=1/2$ for $\U(1)$ breaking, 
$\med{F}=\pi^2/16$ for $\SO(3)\to\SO(2)$, and
$\med{F}=2/3$ for $\SO(4)\to\SO(3)$.

Do walls of light Goldstones behave in a distinctive way?
In the massless limit they do not turn into pulses, unlike in fig.\fig{AdvancingFront},
and keep memory of the intermediate angle.
According to~\cite{2007.04967} this leads to a distinctive GW spectrum.
However, as discussed in section~\ref{postcollisionwall},
the wall-to-pulse transition does not cause energy losses of ultra-relativistic walls.
Energy losses are instead dominated by particle interactions.
Do Goldstones have distinctive interactions?
At low energies, Goldstone scattering cross sections are suppressed by different powers of energy
depending on the group:
\beq
 \Lag_{\rm eff}\ni \frac{1}{f^2} \left\{\begin{array}{l}
 [(\partial_\mu a)^2]^2/2m_s^2 \cr
  f^{abe}f^{cde}a^a(\partial_\mu a^b)\, a^c(\partial_\mu a^d)/6
  \end{array}\right.
  \qquad
\hat{\sigma}(aa\to aa)\sim \frac{\hat{s}}{16\pi f^4}
\left\{\begin{array}{ll}
\hat{s}^2/m_s^4, & \text{Abelian}, \cr
1 & \text{non-Abelian}.
\end{array}\right. 
\eeq
However, above the mass $m_s$ of the radial mode the cross section reduces to the
usual $\hat\sigma \sim \lambda^2/\hat{s}$, where $\lambda$ is the
quartic of the full multiplet, $S=(f+s+ia)/\sqrt2$ in the Abelian case.
The same holds for the other Goldstone interactions:
ultra-relativistic walls made of Goldstones do not have suppressed energy losses.\footnote{For example, Goldstone couplings to fermions are often written as derivative couplings to axial currents,
$(\partial_\mu a)\,(\bar\psi\gamma^\mu\gamma_5\psi)/2f$, equivalent up to the anomaly to
Yukawa couplings $y_{\rm eff}=m_\psi/f$, which look suppressed for light fermions.
Even in KSVZ models, where heavy fermions $Q$ get mass $m_Q=yf/\sqrt2$ from the renormalizable coupling $y\,S\,\bar Q_L Q_R$,
for $ET\gtrsim m_Q^2$ the Goldstone simply interacts through the Yukawa coupling to fundamental fermions,
like the radial mode, as in section~\ref{sec:Yukawa}.
Similarly, the $Q$ loops generate anomalous couplings $a F\tilde F\,\alpha/8\pi f$ to gauge vectors, which allow
Primakoff conversion $ap\to Vp$ via $t$-channel vector exchange, with constant $\hat\sigma\sim\alpha^3/32\pi^2f^2$.
However, when $ET\gtrsim m_Q^2$ the real production $ap\to Q\bar Q p$ is faster by $\sim (4\pi)^2/\alpha$,
with rate $\sim y^2\alpha^2T^3/m_Q^2$  
up to $E\sim m_Q^4/\alpha^2T^3$, above which it becomes the LPM rate of section~\ref{sec:Yukawa}.
}

Finally, topological defects can form where bubbles with different orientations meet,
depending on the topology of the vacuum manifold $G/H$:
domain walls (non-trivial $\pi_0$, for discrete groups) already where 2 bubbles meet;
strings (non-trivial $\pi_1$, as for $\SO(2)\to1$) where 3 or more bubbles meet with orientations that wind around a loop;
monopoles (non-trivial $\pi_2$, as for $\SO(3)\to\SO(2)$) where 4 or more bubbles meet with orientations that wrap a sphere.
No stable domain walls, strings or monopoles form for $\SO(N)\to\SO(N-1)$ with $N\ge4$.
A small explicit breaking that gives mass to the pseudo-Goldstone can attach domain walls to strings.
These possibly problematic objects are known to source GWs (see e.g.~\cite{2111.08750}).




\section{Conclusions}\label{sec:concl}
Gravitational waves from a first-order phase transition are sourced by the anisotropic stress of the expanding bubbles.
For bubble walls with ultra-relativistic $E\gg m$ this stress is localised on thin shells moving at the speed of light,
so that the GW spectrum is determined by the evolution after collisions
of the wall energy per unit area, $\sigma(R)$.
Walls that persist after collision give the `bulk flow' spectrum with an infrared tail $\Omega_{\rm GW}\propto\omega$;
walls that disappear at collision give the `envelope' spectrum with $\Omega_{\rm GW}\propto\omega^3$ at low $\omega$.

So we studied how $\sigma(R)$ evolves after bubble collisions.

Classical effects are negligible:
after `inelastic' collisions the wall spreads into a pulse that conserves its energy (fig.\fig{AdvancingFront}),
and the energy left behind is suppressed by $1/\gamma$.
The decrease of the peak gradient energy density seen in lattice simulations is thereby not an energy loss.

Energy losses are instead dominated by scatterings of the wall quanta on the reheated cosmological plasma.
The problem is analogous to energy losses of ultra-relativistic particles in matter,
an established and non-trivial field~\cite{BetheHeitler,LPM,hep-ph/9802442},
where two kinds of losses are distinguished.
Collisional losses are due to $2\to2$ scatterings on the constituents of the medium, 
and are called ``ionization losses'' in matter, where the targets are atomic electrons.
``Radiative losses'' are due to medium-induced splittings, such as bremsstrahlung and pair conversion.

In the cosmological plasma, hard $2\to2$ scatterings mediated by renormalizable couplings have cross sections $\hat\sigma\sim1/\hat s$
that decrease with energy, while the energy lost per scattering grows with $E$:
the result is a constant $dE/dt\approx -kT^2$, where $k$ is proportional to the squared couplings, as for ionization losses.
These typically stop the walls on time-scales longer than the phase transition.

Faster losses arise from medium-induced splittings of the wall scalar into gauge-charged particles, 
such as $\phi\to\psi\bar\psi$ via a Yukawa coupling or $\phi\to \phi hh^*$ via a quartic coupling to a charged scalar
(see fig.\fig{FeynmanLPMdiagrams}).
They are triggered by soft kicks from $t$-channel exchange of gauge vectors with the plasma.
Unlike electromagnetic showers in matter, in the plasma all splittings with $E\gg T$ are in the Landau--Pomeranchuk--Migdal (LPM) regime 
(for masses $m^2\ll\alpha\sqrt{ET^3}$),
where multiple soft scatterings within the formation time act coherently as one.
Their rate $\sim\alpha T\sqrt{T/E}$ is enhanced by $\alpha\sqrt{E/T}$ compared to hard scatterings,
and can stop the walls within the bubble size.
Since the LPM rate grows as the energy decreases, the subsequent splittings are faster:
the time needed to degrade the energy down to $T$ and isotropize it, ending the GW source,
is dominated by the first splitting~\cite{1301.6102,1405.6318}.

The resulting $\sigma(R)$, summarised in table~\ref{tab:sigmaR} and fig.\fig{Lorentz}, depends on the interaction type and size,
and on whether the interactions degrade or remove the wall quanta.
The GW spectrum thereby provides a window on particle physics.
It ranges between the `envelope' and `bulk flow' limits,
with the $\Omega_{\rm GW}\propto\omega$ infrared tail turning into $\Omega_{\rm GW}\propto\omega^3$ below $1/t_{\rm stop}$,
as shown in fig.\fig{GWSpectrumBubbleCollision}.
Gravitational losses are negligible.

\medskip

We also studied `elastic' collisions, where the bubble intersection returns to the false vacuum:
the walls are decelerated by the same vacuum energy that accelerated them,
turn back and re-collide a few times, radiating at each re-collision a fraction $\sim 30\%$ of their energy
into pulses that behave as in the `inelastic' case.
This mildly modifies the GW spectrum around its peak (fig.\fig{O21ElasticBounces}).
Theories where the transition breaks a global symmetry give a mixture of `elastic' and `inelastic' collisions,
as bubbles nucleated in different directions in field space can overlap in the false vacuum.
Furthermore, the wall energy is shared between radial and Goldstone modes, 
that interact in similar ways with no special  derivative suppression at high energy.
So no distinctive Goldstone GW spectrum arises.

\small \paragraph{Acknowledgements.}
AS thanks Anthropic for a `Claude Team plan for scientists' discount.
This research is not supported by MIUR under PRIN2026.


\footnotesize

\begin{thebibliography}{nnn}\bibitem{Witten:1984rs}
\article{E. Witten}{Phys.Rev.D}{30}{272}{1984}
{\href{https://doi.org/10.1103/PhysRevD.30.272}{Cosmic Separation of Phases}}.


\bibitem{Hogan:1986qda}
\article{C.J. Hogan}{Mon.Not.Roy.Astron.Soc.}{218}{629}{1986}
{Gravitational radiation from cosmological phase transitions}.


\bibitem{Kosowsky:1991ua}
\article{A. Kosowsky, M.S. Turner, R. Watkins}{Phys.Rev.D}{45}{4514}{1992}
{\href{https://doi.org/10.1103/PhysRevD.45.4514}{Gravitational Radiation from Colliding Vacuum Bubbles}}.


\bibitem{2008.09136}
\article[2008.09136]{M.B. Hindmarsh, M. L{\" u}ben, J. Lumma, M. Pauly}{SciPost Phys.Lect.Notes}{24}{1}{2021}
{\href{https://doi.org/10.21468/SciPostPhysLectNotes.24}{Phase transitions in the early universe}}.


\bibitem{2305.02357}
\article[2305.02357]{P. Athron, C. Bal{\' a}zs, A. Fowlie, L. Morris, L. Wu}{Prog.Part.Nucl.Phys.}{135}{104094}{2024}
{\href{https://doi.org/10.1016/j.ppnp.2023.104094}{Cosmological phase transitions: From perturbative particle physics to gravitational waves}}.


\bibitem{1903.09642}
\article[1903.09642]{J. Ellis, M. Lewicki, J.M. No, V. Vaskonen}{JCAP}{06}{024}{2019}
{\href{https://doi.org/10.1088/1475-7516/2019/06/024}{Gravitational wave energy budget in strongly supercooled phase transitions}}.


\bibitem{2307.04694}
\article[2307.04694]{A. Salvio}{JCAP}{12}{046}{2023}
{\href{https://doi.org/10.1088/1475-7516/2023/12/046}{Supercooling in Radiative Symmetry Breaking: Theory Extensions, Gravitational Wave Detection and Primordial Black Holes}}.


\bibitem{1604.05035}
\article[1604.05035]{R. Jinno, M. Takimoto}{Phys.Rev.D}{95}{015020}{2017}
{\href{https://doi.org/10.1103/PhysRevD.95.015020}{Probing a classically conformal B-L model with gravitational waves}}.


\bibitem{1704.01034}
\article[1704.01034]{L. Marzola, A. Racioppi, V. Vaskonen}{Eur.Phys.J.C}{77}{484}{2017}
{\href{https://doi.org/10.1140/epjc/s10052-017-4996-1}{Phase transition and gravitational wave phenomenology of scalar conformal extensions of the Standard Model}}.


\bibitem{1704.04955}
\article[1704.04955]{S. Iso, P.D. Serpico, K. Shimada}{Phys.Rev.Lett.}{119}{141301}{2017}
{\href{https://doi.org/10.1103/PhysRevLett.119.141301}{QCD-Electroweak First-Order Phase Transition in a Supercooled Universe}}.


\bibitem{1805.01473}
\article[1805.01473]{T. Hambye, A. Strumia, D. Teresi}{JHEP}{08}{188}{2018}
{\href{https://doi.org/10.1007/JHEP08(2018)188}{Super-cool Dark Matter}}.


\bibitem{1809.11129}
\article[1809.11129]{T. Prokopec, J. Rezacek, B. Swiezewska}{JCAP}{02}{009}{2019}
{\href{https://doi.org/10.1088/1475-7516/2019/02/009}{Gravitational waves from conformal symmetry breaking}}.


\bibitem{BodekerMoore}
\article[0903.4099]{D. Bodeker, G.D. Moore}{JCAP}{05}{009}{2009}
{\href{https://doi.org/10.1088/1475-7516/2009/05/009}{Can electroweak bubble walls run away?}}


\bibitem{1703.08215}
\article[1703.08215]{D. Bodeker, G.D. Moore}{JCAP}{05}{025}{2017}
{\href{https://doi.org/10.1088/1475-7516/2017/05/025}{Electroweak Bubble Wall Speed Limit}}


\bibitem{1906.02588}
\article[1906.02588]{R. Jinno, T. Konstandin, M. Takimoto}{JCAP}{09}{035}{2019}
{\href{https://doi.org/10.1088/1475-7516/2019/09/035}{Relativistic bubble collisions—a closer look}}.


\bibitem{1912.00997}
\article[1912.00997]{M. Lewicki, V. Vaskonen}{Phys.Dark Univ.}{30}{100672}{2020}
{\href{https://doi.org/10.1016/j.dark.2020.100672}{On bubble collisions in strongly supercooled phase transitions}}.


\bibitem{2007.04967}
\article[2007.04967]{M. Lewicki, V. Vaskonen}{Eur.Phys.J.C}{80}{1003}{2020}
{\href{https://doi.org/10.1140/epjc/s10052-020-08589-1}{Gravitational wave spectra from strongly supercooled phase transitions}}.


\bibitem{GiblinHuiLimYang}
\article[1005.3493]{G.D. Moore, L. Hui, E.A. Lim, I-S. Yang}{Phys.Rev.D}{82}{045019}{2010}
{\href{https://doi.org/10.1103/PhysRevD.82.045019}{How to Run Through Walls: Dynamics of Bubble and Soliton Collisions}}


\bibitem{AhlqvistEckerleGreene}
\article[1310.6069]{P. Ahlqvist, K. Eckerle, B. Greene}{JHEP}{03}{031}{2015}
{\href{https://doi.org/10.1007/JHEP03(2015)031}{Bubble Universe Dynamics After Free Passage}}


\bibitem{2012.07826}
\article[2012.07826]{M. Lewicki, V. Vaskonen}{Eur.Phys.J.C}{81}{437}{2021}
{\href{https://doi.org/10.1140/epjc/s10052-021-09232-3}{Gravitational waves from colliding vacuum bubbles in gauge theories}}.


\bibitem{2208.11697}
\article[2208.11697]{M. Lewicki, V. Vaskonen}{Eur.Phys.J.C}{83}{109}{2023}
{\href{https://doi.org/10.1140/epjc/s10052-023-11241-3}{Gravitational waves from bubble collisions and fluid motion in strongly supercooled phase transitions}}.


\bibitem{1707.03111}
\article[1707.03111]{R. Jinno, M. Takimoto}{JCAP}{01}{060}{2019}
{\href{https://doi.org/10.1088/1475-7516/2019/01/060}{Gravitational waves from bubble dynamics: Beyond the Envelope}}.


\bibitem{2306.15555}
\article[2306.15555]{I. Baldes, M. Dichtl, Y. Gouttenoire, F. Sala}{Phys.Rev.Lett.}{134}{6}{2025}
{\href{https://doi.org/10.1103/PhysRevLett.134.061001}{Ultrahigh-Energy Particle Collisions and Heavy Dark Matter at Phase Transitions}}


\bibitem{2403.05615}
\article[2403.05615]{I. Baldes, M. Dichtl, Y. Gouttenoire, F. Sala}{JHEP}{07}{231}{2024}
{\href{https://doi.org/10.1007/JHEP07(2024)231}{Particle shells from relativistic bubble walls}}


\bibitem{BetheHeitler}
\article{H. Bethe, W. Heitler}{Proc.Roy.Soc.Lond.A}{146}{83}{1934}
{\href{https://doi.org/10.1098/rspa.1934.0140}{On the Stopping of fast particles and on the creation of positive electrons}}.


\bibitem{LPM}
\article{L.D. Landau, I. Pomeranchuk}{Dokl.Akad.Nauk Ser.Fiz.}{92}{535}{1953}
{Limits of applicability of the theory of bremsstrahlung electrons and pair production at high-energies}.
\article{L.D. Landau, I. Pomeranchuk}{Dokl.Akad.Nauk Ser.Fiz.}{92}{735}{1953}
{Electron cascade process at very high-energies}.
\article{A.B. Migdal}{Phys.Rev.}{103}{1811}{1956}
{\href{https://doi.org/10.1103/PhysRev.103.1811}{Bremsstrahlung and Pair Production at High Energies in Condensed Media}}.


\bibitem{hep-ph/9802442}
\article[hep-ph/9802442]{S. Klein}{Rev.Mod.Phys.}{71}{1501}{1999}
{\href{https://doi.org/10.1103/RevModPhys.71.1501}{Suppression of Bremsstrahlung and pair production due to environmental factors}}.


\bibitem{BDMPS}
\article[hep-ph/9607355]{R. Baier, Y.L. Dokshitzer, A.H. Mueller, S. Peign\'e, D. Schiff}{Nucl.Phys.B}{483}{291}{1997}
{\href{https://doi.org/10.1016/S0550-3213(96)00553-6}{Radiative energy loss of high-energy quarks and gluons in a finite volume quark - gluon plasma}}.
\article[hep-ph/9608322]{R. Baier, Y.L. Dokshitzer, A.H. Mueller, S. Peign\'e, D. Schiff}{Nucl.Phys.B}{484}{265}{1997}
{\href{https://doi.org/10.1016/S0550-3213(96)00581-0}{Radiative energy loss and p(T) broadening of high-energy partons in nuclei}}.
\article[hep-ph/9607440]{B.G. Zakharov}{JETP Lett.}{63}{952}{1996}
{\href{https://doi.org/10.1134/1.567126}{Fully quantum treatment of the Landau-Pomeranchuk-Migdal effect in QED and QCD}}.


\bibitem{AMY}
\article[hep-ph/9804224]{P. Aurenche, F. Gelis, R. Kobes, H. Zaraket}{Phys.Rev.D}{58}{085003}{1998}
{\href{https://doi.org/10.1103/PhysRevD.58.085003}{Bremsstrahlung and photon production in thermal QCD}}.
\article[hep-ph/0111107]{P.B. Arnold, G.D. Moore, L.G. Yaffe}{JHEP}{12}{009}{2001}
{\href{https://doi.org/10.1088/1126-6708/2001/12/009}{Photon emission from quark gluon plasma: Complete leading order results}}.
\article[hep-ph/0204343]{P.B. Arnold, G.D. Moore, L.G. Yaffe}{JHEP}{06}{030}{2002}
{\href{https://doi.org/10.1088/1126-6708/2002/06/030}{Photon and gluon emission in relativistic plasmas}}.


\bibitem{astro-ph/9211004}
\article[astro-ph/9211004]{A. Kosowsky, M.S. Turner}{Phys.Rev.D}{47}{4372}{1993}
{\href{https://doi.org/10.1103/PhysRevD.47.4372}{Gravitational radiation from colliding vacuum bubbles: envelope approximation to many bubble collisions}}.


\bibitem{0806.1828}
\article[0806.1828]{S.J. Huber, T. Konstandin}{JCAP}{09}{022}{2008}
{\href{https://doi.org/10.1088/1475-7516/2008/09/022}{Gravitational Wave Production by Collisions: More Bubbles}}.


\bibitem{1712.06869}
\article[1712.06869]{T. Konstandin}{JCAP}{03}{047}{2018}
{\href{https://doi.org/10.1088/1475-7516/2018/03/047}{Gravitational radiation from a bulk flow model}}.


\bibitem{2308.13070}
\article[2308.13070]{H. Mansour, B. Shakya}{Phys.Rev.D}{111}{023520}{2025}
{\href{https://doi.org/10.1103/PhysRevD.111.023520}{Particle production from phase transition bubbles}}.


\bibitem{2511.15687}
\article[2511.15687]{M. Lewicki, V. Vaskonen}{Phys.Dark Univ.}{53}{102425}{2026}
{\href{https://doi.org/10.1016/j.dark.2026.102425}{Impact of cosmic expansion on gravitational wave spectra from strongly supercooled first-order phase transitions}}.


\bibitem{Weinberg:1965nx}
\article{S. Weinberg}{Phys.Rev.}{140}{B516}{1965}
{\href{https://doi.org/10.1103/PhysRev.140.B516}{Infrared photons and gravitons}}.


\bibitem{1605.01403}
\article[1605.01403]{R. Jinno, M. Takimoto}{Phys.Rev.D}{95}{024009}{2017}
{\href{https://doi.org/10.1103/PhysRevD.95.024009}{Gravitational waves from bubble collisions: An analytic derivation}}.


\bibitem{astro-ph/9310044}
\article[astro-ph/9310044]{M. Kamionkowski, A. Kosowsky, M.S. Turner}{Phys.Rev.D}{49}{2837}{1994}
{\href{https://doi.org/10.1103/PhysRevD.49.2837}{Gravitational radiation from first order phase transitions}}.


\bibitem{LPMexp}
\article{SLAC E-146 Collaboration}{Phys.Rev.Lett.}{75}{1949}{1995}
{\href{https://doi.org/10.1103/PhysRevLett.75.1949}{An Accurate measurement of the Landau-Pomeranchuk-Migdal effect}}.


\bibitem{1312.3097}
\article[1312.3097]{K. Harigaya, K. Mukaida}{JHEP}{05}{006}{2014}
{\href{https://doi.org/10.1007/JHEP05(2014)006}{Thermalization after/during Reheating}}


\bibitem{Weldon:1982aq}
\article{H.A. Weldon}{Phys.Rev.D}{26}{1394}{1982}
{\href{https://doi.org/10.1103/PhysRevD.26.1394}{Covariant Calculations at Finite Temperature: The Relativistic Plasma}}.


\bibitem{0810.1026}
\article[0810.1026]{P.B. Arnold, W. Xiao}{Phys.Rev.D}{78}{125008}{2008}
{\href{https://doi.org/10.1103/PhysRevD.78.125008}{High-energy jet quenching in weakly-coupled quark-gluon plasmas}}.


\bibitem{1905.00899}
\article[1905.00899]{R. Jinno, H. Seong, M. Takimoto, C.M. Um}{JCAP}{10}{033}{2019}
{\href{https://doi.org/10.1088/1475-7516/2019/10/033}{Gravitational waves from first-order phase transitions: ultra-supercooled transitions and the fate of relativistic shocks}}


\bibitem{2308.12916}
\article[2308.12916]{R. Sharma, J. Dahl, A. Brandenburg, M. Hindmarsh}{JCAP}{12}{042}{2023}
{\href{https://doi.org/10.1088/1475-7516/2023/12/042}{Shallow relic gravitational wave spectrum with acoustic peak}}.


\bibitem{2308.12943}
\article[2308.12943]{A. Roper Pol, S. Procacci, C. Caprini}{Phys.Rev.D}{109}{063531}{2024}
{\href{https://doi.org/10.1103/PhysRevD.109.063531}{Characterization of the gravitational wave spectrum from sound waves within the sound shell model}}.


\bibitem{2111.08750}
\article[2111.08750]{D.I. Dunsky, A. Ghoshal, H. Murayama, Y. Sakakihara, G. White}{Phys.Rev.D}{106}{075030}{2022}
{\href{https://doi.org/10.1103/PhysRevD.106.075030}{GUTs, hybrid topological defects, and gravitational waves}}.


\bibitem{1301.6102}
\article[1301.6102]{J.-P. Blaizot, E. Iancu, Y. Mehtar-Tani}{Phys.Rev.Lett.}{111}{052001}{2013}
{\href{https://doi.org/10.1103/PhysRevLett.111.052001}{Medium-induced QCD cascade: democratic branching and wave turbulence}}


\bibitem{1405.6318}
\article[1405.6318]{A. Kurkela, E. Lu}{Phys.Rev.Lett.}{113}{182301}{2014}
{\href{https://doi.org/10.1103/PhysRevLett.113.182301}{Approach to Equilibrium in Weakly Coupled Non-Abelian Plasmas}}


\end{thebibliography}
\end{document}